\documentclass{aa}  

\usepackage{graphicx}
\usepackage{txfonts}
\usepackage[]{hyperref} 
\usepackage{booktabs} 
\usepackage{array} 
\usepackage{paralist} 
\usepackage{verbatim} 
\usepackage{ulem}
\usepackage{xspace}
\usepackage{siunitx}
\usepackage{amssymb}
\usepackage{xcolor}
\usepackage{float}
\usepackage{comment}

\usepackage{natbib}
\bibpunct{(}{)}{;}{a}{}{,} 
\graphicspath{{./img/}}

\hypersetup{
    colorlinks=true,
    linkcolor=blue,
    filecolor=blue,
    urlcolor=blue,
    citecolor=blue
}

\makeatletter
\renewcommand*\aa@pageof{, page \thepage{} of \pageref*{LastPage}}
\makeatother

\addto\extrasenglish{%
}

\newcommand{\sysrem}{SYSREM\xspace}
\newcommand{\petitradtrans}{petitRADTRANS\xspace}
\newcommand{\crires}{CRIRES+\xspace}
\newcommand{\citepeg}{\citep[e.g.][]}

\newcommand{\Mjupdot}{$M_J$}
\newcommand{\Rjup}{$R_J$ }

\newcommand{\about}{\ensuremath{{\sim}}}
\newcommand{\Ss}{$\mathcal{S}$}
\newcommand{\Sss}{$\mathcal{S}$ }
\begin{document}

   \title{Optimising spectroscopic observations of transiting exoplanets}


   \author{
            Linn Boldt-Christmas \inst{1} \and
            Fabio Lesjak \inst{2} \and 
            Ansgar Wehrhahn \inst{1} \and
            Nikolai Piskunov \inst{1} \and
            Adam D. Rains \inst{1}  \and
            Lisa Nortmann \inst{2} \and
            Oleg Kochukhov \inst{1}
          }

   \institute{
            Observational Astrophysics, Department of Physics and Astronomy, Uppsala University, Sweden \and
            Institut f\"ur Astrophysik, Georg-August-Universit\"at, Friedrich-Hund-Platz 1, 37077 G\"ottingen, Germany
             }

   \date{Received July 10, 2023; accepted December 13, 2023}

 
  \abstract
   {When observing the atmospheres of transiting exoplanets using high-resolution spectroscopy, one aims to detect well-resolved spectral features with high signal-to-noise ratios (SNR) as is possible today with modern
spectrographs. However, obtaining such high-quality observations comes with a trade-off: a lower cadence of fewer, longer exposures across the transit collects more photons thanks to reduced overheads, enhancing the SNR of each observation, while a higher cadence of several, shorter exposures minimises spectral feature smearing due to the
continuously changing radial velocity of the planet.}  
   {Considering that maximising SNR and minimising smearing are both beneficial to analysis, there is a need to establish what the optimal compromise is between the two for a given target. In this work, we aim to establish where this compromise lies for a typical exoplanet transit observation in order to benefit future data collection and subsequent interpretation.}
   {In this work, we model real transit events based on targets as they would be observed with VLT/\crires at Paranal Observatory, Chile. Creating four hypothetical scenarios, we simulate each set of transmission spectra across 100 realisations of the same transit event in order to vary the time resolution only. We remove telluric and stellar lines from these data sets using the \sysrem algorithm and analyse them through cross-correlation with model templates, measuring how successfully each time resolution and case detected the planetary signal and exploring how the results vary.}
   {We demonstrate that there is a continuous change
in the significance of the cross-correlation detection based on
the trade-off between high and low time resolutions, and that averaged over a
large number of realisations, the function of this significance has clear
maxima. The strength and location of this function's maxima varies depending on e.g. planet system parameters, instrumentation, and number of removal iterations. We discuss why observers should therefore take several factors into account, using a strategy akin to the `exposure triangle' employed in traditional photography where a balance must be struck by considering the full context of the observation. Our method is robust and may be employed by observers to estimate best observational strategies for other targets.}
   {}


   \maketitle
%

\section{Introduction}
    \label{section:Introduction}

    As exoplanets orbit their host stars, the chemical and physical structures of their atmospheres can be studied using spectroscopic observations. New exoplanet candidates are continually being discovered by transit-searching surveys, including both ground-based surveys like WASP \citep{pollacco_wasp_2006}, HATNet \citep{bakos_wide-field_2004}, HATSouth \citep{bakos_hatsouth_2013}, KELT \citep{pepper_kilodegree_2007}, TRAPPIST \citep{jehin_trappist_2011}, and MASCARA \citep{talens_mascara-1_2017}, as well as space-based surveys like Kepler \citep{borucki_kepler_2010}, K2 \citep{howell_k2_2014}, and TESS \citep{ricker_transiting_2015}. Together, such surveys have resulted in over 5,500 confirmed exoplanets on record today, with over 4,500 of those being discovered in the last ten years.\footnote{NASA Exoplanet Archive: \url{https://exoplanetarchive.ipac.caltech.edu/}} Thanks to this progress, alongside our increasing ability to study these planets in detail with new instruments and improvements in our methods,  the relatively new field of exoplanet atmosphere characterisation is rapidly maturing. 


    Spectroscopic characterisation can be done through transmission observations of exoplanets' nightsides or through emission and/or reflection observations of their daysides. Most confirmed exoplanetary systems are observed edge-on meaning that planets can be seen regularly transiting across the face of their host stars (a primary eclipse), during which we observe the dusk-dawn terminator on their nightsides and collect light that has been transmitted through the upper layers of the planet atmosphere. Beyond this eclipse, the planets can also be studied as they continue along their orbit and are illuminated by their host star, with this dayside reflecting light from the host star together with whatever emission features come from the irradiated planet. This type of observation is ideally done immediately prior to or following the planet disappearing behind the host (a secondary eclipse), but has a much larger window of opportunity in time and has the advantage that it can be done for non-transiting systems too. 
    
    As such, transmission studies face unique challenges in its constraints. The two branches of transit spectroscopy – low-resolution from space and high-resolution from the ground – both provide impressive results thanks to their specific and complementary strengths. Low spectral resolution has the drawback that one cannot distinguish individual spectral lines, but space-based observations benefit from the lack of telluric contamination and flux-calibrated spectra with high photometric precision across a wide spectral range. This means that the contrast between wavelengths with strong absorption (where the exoplanet atmosphere is opaque) and those with little absorption is diminished with low-resolution, but the remaining effect can be detected with high photometric precision. This is possible from space, and characterisation has been regularly achieved with space-based telescopes such as the Hubble Space Telescope (as \citet{seager_theoretical_2000} predicted would be possible) for over twenty years now \citepeg{charbonneau_detection_2002,vidal-madjar_detection_2004,swain_presence_2008,pont_detection_2008,sing_hubble_2008,sing_hubble_2011,kreidberg_clouds_2014,stevenson_quantifying_2016,macdonald_hd_2017,kreidberg_water_2018,benneke_water_2019,zhou_hubble_2022} and more recently, by JWST \citepeg{rustamkulov_jwst_2023,jwst_transiting_exoplanet_community_early_release_science_team_identification_2023}. 

Comparatively, the mirror size and associated instrumentation required for high spectral resolution observations across a large wavelength range poses an engineering challenge that is currently unachievable (in part fiscally) to do from space, and as such, this is restrained to ground-based observations. High spectral resolution allows lines to be more easily separated, and for ground-based observations where spectral resolution can be high across a large wavelength range, the high contrast between the effective blocking area of the planet in different wavelengths provides the ability to look between the cores of strong telluric absorption lines that dominate most of the near infrared. High-resolution ground-based results have been consistently obtained for over a decade \citepeg{redfield_sodium_2008,snellen_ground-based_2008,snellen_orbital_2010,bean_ground-based_2010,astudillo-defru_ground-based_2013,di_gloria_using_2015,allart_search_2017,brogi_exoplanet_2018,diamond-lowe_ground-based_2018,merritt_inventory_2021,nikolov_ground-based_2021, maimone_detecting_2022,maguire_high-resolution_2023}, and more recently, even in collaboration with space-based results \citepeg{boucher_co_2023,spyratos_precise_2023}. 

Together, these efforts are contributing to an expanding catalogue of confirmed exoplanets and an evolving understanding of their composition, revealing the diversity of exoplanetary systems' architectures. With the advent of highly stable spectrographs on the largest ground-based telescopes combining wide wavelength range with high spectral resolution, such as HARPS \citep{mayor_setting_2003}, IGRINS \citep{park_design_2014}, SPIRou \citep{artigau_spirou_2014}, MAROON-X \citep{seifahrt_-sky_2020}, ESPRESSO \citep{pepe_espresso_2021}, and CRIRES+ \citep{dorn_crires_2023}, the number of exoplanets well-characterised with high-resolution spectroscopy is steadily increasing. Cross-correlation analysis with realistic templates initially made it possible to perform analyses of hot Jupiter atmospheres' compositions and basic dynamics \citep{brogi_rotation_2016,hoeijmakers_atomic_2018}, and today, the quality of observations and the advances of the analysis techniques (for many instruments, including benefits from adaptive optics systems) have reached the point when one can even speculate about more detailed meteorological effects. This is possible as we are starting to identify horizontal and vertical stratification on exoplanets as well as dynamic weather patterns as inferred from the shape of individual lines, which are resolvable with high-quality, high-resolution spectroscopy \citepeg{ehrenreich_nightside_2020,pino_gaps_2022,prinoth_titanium_2022,gandhi_retrieval_2023,yan_crires_2023}.
    
    This ability to identify spectral features in detail is not only enabling more detailed characterisation, but it is also relevant as studies indicate that occurrence rates of exoplanets are high around cooler stars such as FGK stars \citep{kunimoto_searching_2020} and M-dwarfs \citep{hardegree-ullman_kepler_2019,kanodia_mass-radius_2019}. M-dwarfs also see an overabundance of low-mass planets with shorter periods in particular \citep{sabotta_carmenes_2021}, meaning they are especially suitable for transit studies. As the photospheres of cooler stars allow complex molecular species to exist, these colder planet-hosting stars suffer from more chemically complicated stellar atmospheres whose absorption features flood the continuum, creating greater challenges for disentangling the planetary and stellar spectra \citep{wakeford_disentangling_2019}.
    
    Great efforts are going into extending the field's successes so far in characterising gas giants to the domain of low-mass planets in potentially habitable zones. This task remains a challenge for (at least) observations and so in this paper we explore one important question of high-resolution transit spectroscopy: how to obtain the best possible data by considering the trade-off between signal-to-noise ratio (SNR) and the time resolution of observations. 
    
    During the observation of a transit, there are two possible strategies that may be employed: exposures can either be taken at a lower cadence of fewer, longer exposures or at a higher cadence of several, shorter exposures. Across the transit, longer exposures collect more photons thanks to reduced overheads, which enhances the SNR of each exposure. Meanwhile, shorter exposures minimise the effect where spectral features get smeared due to the planet's radial velocity that is continuously changing across the single exposure. Considering that both maximising the SNR per exposure and minimising the effect of smearing will benefit analysis, there is a need to establish what the optimal compromise is between the two for a given target.

      In this paper, we explore this balance by investigating what range of time resolutions result in the strongest possible planetary signal with traditional cross-correlation analysis (as a measure of what can be considered an `optimal' time resolution). We demonstrate that there is a continuous change in the significance of the cross-correlation detection based on the trade-off between a small number of higher SNR exposures and a large number of lower SNR exposures of a transiting target, and that averaged over a large number of realisations, the function of this significance has clear maxima between the two. By testing different case studies, we find that the location of this maxima depends on a number of factors including target/system parameters. With this work, we are also presenting a robust method that may be employed to determine the ideal time resolution for observing other given targets using simulated spectra. Using the methods presented here may help observers plan what exposure cadence to use prior to taking their data, therefore optimising their own transit spectroscopy observations.
    
In Section \ref{section:Method}, we will describe how we explore this question using model spectra (i.e. simulated observations). We describe how these spectra are generated to create a set of data that is representative of observations possible to obtain with \crires while still maintaining full control of external variables for the purpose of comparability and repeatability. In Section \ref{section:Analysis}, we describe how our simulated observations are then analysed (following the same methodology that would be used on real data), removing the stellar signal and the telluric contamination with the commonly used \sysrem method before performing cross-correlation analysis on the remaining planetary signal. We present these results in Section \ref{section:Results}, followed by some further discussion regarding other considerations in Section \ref{section:Discussion}. Final conclusions and recommendations are presented in Section \ref{section:Conclusions}.

\section{Method}
\label{section:Method}

 \subsection{Time resolution optimisation problem}
        \label{subsection:OptimisationProblem}
   
A transmission spectrum observed from Earth consists of three components: (i) stellar spectral lines from the exoplanet host star, (ii) telluric spectral lines from the Earth's atmosphere, and (iii) the minor contribution of spectral lines from the exoplanet atmosphere. Importantly, the spectral and telluric lines are relatively stationary across the transit in comparison to the lines of the orbiting exoplanet; as the exoplanet moves towards and then away from the observer, its lines will vary from being blueshifted at the beginning of its transit and redshifted towards the end. In analysis, this difference in Doppler shifts between lines of different origins is exploited to help distinguish the planetary component from the stellar and telluric component (see subsection \ref{subsection:SYSREM}).

    The planetary component is only found in the spectra of exposures taken during the actual eclipse of the host star by the moving planet. This is limited in time, and so it sets a firm limitation on the possible maximum SNRs of the data collected with a given instrument. In practice, observations are also taken before and after the transit to assess the stability of the instrument and to extend the baseline for following data trends such as the strong telluric features, but this data does not contain any planetary signal.
    
    Consider the extreme case where a transit observation consists of a single exposure over an entire transit. In this case, light collection and thus SNR is maximised (much like taking a long-exposure photograph in the dark on Earth) but all information about radial velocity shifts of the planet is lost as one cannot see the Doppler shift of the planetary component. This would render the data useless because subsequent analysis will not be able to tell apart planetary, telluric, and stellar features. The total transit exposure must therefore be split up into some number of subdivisions in order to track this shift, even if this results in some loss of SNR. 
    
    On the other end of extremes, in the case of too many subdivisions, there must also be a lower limit of SNR below which the cross-correlation signal will be dominated by the noise. This is because if one ignores readout noise and exposure overheads, the total SNR is (approximately) preserved while subdividing exposures; but in reality, what sets the limit is the background and the readout noise, and both grow linearly in combined data with the number of subexposures.

\begin{figure}
    \centering
    \includegraphics[width=\columnwidth]{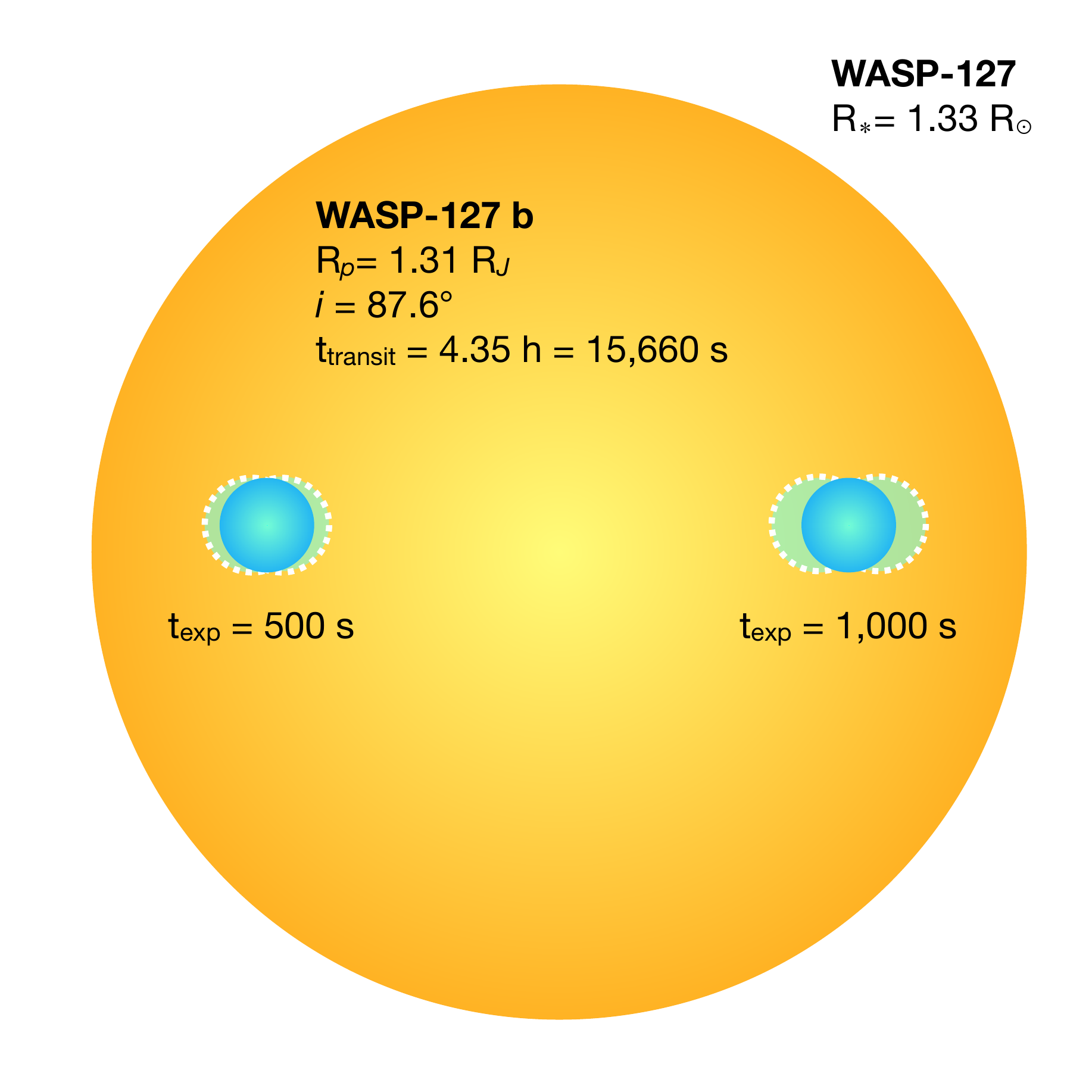}
    \caption{Schematic demonstrating how the smearing effect arises due to increased exposure length. Using an illustration of exoplanet WASP-127 b and its host star (to scale) as an example, consider the target's transit time $t_{\mathrm{transit}}$ = 4.35 hours = 15,660 seconds. A shorter exposure time of $t_{\mathrm{exp}}$ = 500 seconds (left) covers approx. 3\% of the total transit time, versus a longer exposure time of $t_{\mathrm{exp}}$ = 1,000 seconds (right) which covers approx. 6\%. In this schematic, the location of the planet at the beginning and end of an exposure of length $t_{\mathrm{exp}}$ is shown by the shaded, dotted regions. The relative distance between these regions illustrates how much the target moves across a single exposure of the given time. This movement is greater for the longer exposure and thus will give rise to a larger smearing effect.}
    \label{fig:smearing_schematic}
\end{figure}

    Establishing an appropriate number of exposures across the transit is not straightforward. Observing a transit at a lower time resolution, meaning we take a smaller number of longer exposures, increases total light collection per exposure as less time is lost to overheads, which increases SNR. However, this also results in an effect where resolved spectral features captured over a longer period are `smeared' during the exposure due to the changing radial velocity of the moving target. A schematic illustrating how this smearing effect arises is shown in Figure \ref{fig:smearing_schematic}.
    
    This smearing effect makes analysis and the identification of spectral lines more challenging or even impossible, especially when trying to retain the true line profile shape for more detailed atmospheric characterisation. Conversely, at a higher time resolution, meaning a larger number of shorter exposures, there is less smearing of spectral features but more time is lost in overheads between exposures (decreasing total exposure time) which gives a lower total light collection and lower SNR on the single exposure. As retaining line profiles with a high time resolution and obtaining high SNR with a low time resolution are both beneficial for analysis, there is a need to strike a balance between the two. 

    \subsection{Generating model \crires spectra}\label{subsection:SimsObs}

Our objective is to analyse observational data of different time resolutions to determine  where the balance between high and low exposure cadence lies. In order to analyse an ideal scenario with all uncertainties controlled, we produced simulated model spectra to conduct our analysis on. This is because if one was to attempt an investigation of our time resolution optimisation problem using real observational data only, the immediate obstacle would be that one cannot test different time resolutions simultaneously. Consequently, one would have to settle for observations of the same target using different time resolutions taken across different nights – and all would have different weather conditions, visibility, lunar contamination, etc. As such, using simulated spectra ensures that all other parameters unrelated to our study are fully controlled, and allows us to play with the level of the planetary signal for different targets to follow detection deterioration.

Since the results are instrument- and target-specific, the model spectra were designed to be representative of the type and quality of data that can be delivered by \crires at the VLT at Paranal Observatory in Chile. \crires is the upgrade project of the previous CRIRES instrument (CRyogenic InfraRed Echelle Spectrograph, in operation until 2014) that was recently completed with first science observations conducted in October 2021. \crires is a cross-dispersed spectrograph with a nominal spectral resolution of up to approximately $R$ = 100,000 across wavelengths from \SIrange{0.95}{5.3}{\micro\meter} (YJHKLM bands), covering the near-infrared region and partial mid-infrared region.  This range is vital for resolving the spectral lines of several molecular species (such as CO, CO$_2$, H$_2$O, NH$_3$, and CH$_4$), and notably, \crires is rare in its ability to cover the domain of \SIrange{3.0}{5.2}{\micro\meter} at high-resolution \citep{dorn_crires_2023}. The range of \crires is also highly complementary with the spectroscopic capabilities of the recently launched JWST, namely the NIRSpec instrument which covers \SIrange{0.6}{5.3}{\micro\meter} \citep{birkmann_near-infrared_2022,jakobsen_near-infrared_2022}.

Each simulated data set corresponds to a real transit event of a hypothetical target planet as visible from Paranal, and comes with realistic time-dependent local conditions such as air mass variation, barycentric velocities for each exposure, and the telluric absorption computed for the zenith distance of the target at the moment of observations. Target and system parameters were taken from a selected case study target, WASP-127 b. This target is a puffy gas giant of radius 1.31 \Rjup and mass 0.165 \Mjupdot, found at a semi-major axis distance of 0.0484 au from its G5 host star with an orbital period of 4.18 days and a transit duration of 4.35 hours \citep{seidel_hot_2020} meaning it can reasonably be studied using a single transit. 

The transit observations were simulated using a custom script\footnote{\url{https://github.com/fabioles/Eclspec}} that generates spectral data by following the path of the light from the stellar surface to the spectrograph detector. All simulation parameters can be found in Table \ref{tab:SimsParameters}. 
\renewcommand{\arraystretch}{1.2}
\begin{table}\small
    \centering
        \caption{Table of all parameters used for generating simulated spectra of reference case study, a fiducial planet (based on exoplanet WASP-127 b) as it would be observed with VLT/\crires.}
    \begin{tabular}{ lrr } 
     \toprule
    \textbf{Star (WASP-127)} & & \\
        \midrule
       Parameter & Units & Value \\
     \midrule
      Mass$^a$ & $M_* [M_\odot]$ & 0.950 $\pm$ 0.020 \\
      Radius$^a$ & $R_* [R_\odot]$ & 1.333 $\pm$ 0.027\\
      Spectral class$^b$ & & G5 \\
      Magnitude (K-band)$^c$ & $m_{K_s}$ & 8.641 $\pm$ 0.019 \\ 
      Effective temperature$^d$ & $T_\mathrm{eff}$ [K] & 5828.010 $^{+107.743} _{-104.525}$ \\
      Metallicity$^d$ & [M/H] & -0.193 $\pm$ 0.014 \\
      Log. gravity (log $g$)$^d$ & log$_{10}$ [cm/s$^2$] & 4.200 $^{+0.0771} _{-0.064}$ \\
      Systemic radial velocity$^e$ & [km/s] & -8.248 $\pm$ 0.892 \\ 
    \midrule
    \midrule
    \textbf{Planet (WASP-127 b)} & & \\
    \midrule
    Parameter & Symbol & Value \\
     \midrule
     Mass$^a$ & $M_p [M_\mathrm{J}]$  & 0.1647 $^{+0.0214} _{-0.0172}$ \\
     Radius$^a$ & $R_p [R_\mathrm{J}]$ & 1.311 $^{+0.025} _{-0.029}$ \\
     Semi-major axis$^a$ & $a$ [au] & 0.0484 $^{+0.0014} _{-0.0010}$ \\
     Orbital period$^a$ & $P$ [days] & 4.1780620 $^{+0.0000009}_{-0.0000005}$\\ 
     Orbital eccentricity$^a$ & $e$ & 0 \\
     Orbital inclination$^a$ & $i$ (deg) & 87.84 $^{+0.36}_{-0.33}$  \\
     Radial velocity amplitude$^a$ & $K$ (m/s) & 22 $^{+3}_{-2}$ \\
     Transit duration$^a$ & $t_\mathrm{t}$ [hours] & 4.3529 $^{+0.0084}_{-0.0139}$\\
     Transit depth$^a$ & $\delta$  $[\%]$  & 1.021 $^{+0.005}_{-0.029}$  \\
     Assumed eq. temperature$^a$ & $T_\mathrm{eq}$ [K] & 1400 $\pm$ 24 \\
     \midrule
     \midrule
          \textbf{Observing (VLT/\crires)} & & \\
        \midrule
       Parameter & Symbol & Value \\
     \midrule
    Wavelength setting$^f$ &  & \texttt{K2148} \\
   Wavelength coverage$^g$ &  {\textmu}m & 2.007 - 2.491 \\
    No. of echelle orders$^f$  & & 6 \\
    Air mass variation$^h$ &  & 1.58 - 1.06 - 1.99 \\
    Seeing$^i$ & & 0.7 \\
    Readout time$^g$  & $t_{RO}$ [s] & 14 \\

     \bottomrule
      & & \\

    \end{tabular}
    \textbf{References:} $^a$ \citet{seidel_hot_2020},
    $^b$ \citet{lam_dense_2017},
    $^c$ 2MASS All-Sky Catalog \citep{cutri_vizier_2003}, $^d$ TESS Input Catalog \\ \citep{stassun_revised_2019}, $^e$ Gaia DR2 \citep{gaia_collaboration_gaia_2018}, \\ $^f$ ESO/CRIRES+ User Manual, $^g$ \citet{dorn_crires_2023}, $^h$ Calculated using \texttt{pyasl.airmassPP} from PyAstronomy, $^i$ Average Paranal seeing in 2019 according to data from the ESO Differential Image Motion Monitor \citep{kornilov_combined_2007}
    \label{tab:SimsParameters}
\end{table}

Our spectral simulator first sets up the Keplerian orbits of both the planet and host star around the barycentre according to the catalogued ephemerides, and the orbits are used to determine positions and velocities of the two bodies for each exposure. An appropriate stellar template spectrum $I_\star$ is then selected from the PHOENIX spectral library \citep{husser_new_2013}. For the planetary spectrum, we use a reference exoplanet atmosphere transmission spectrum that has been computed using the radiative transfer package \petitradtrans \citep{molliere_petitradtrans_2019} and the HITEMP line list for CO and H$_2$O \citep{rothman_hitemp_2010}, modelling a theoretical atmosphere based on parameters for a generic hot Jupiter atmosphere \citep{bouchy_elodie_2005,addison_minerva-australis_2019}. This planetary spectrum is then imprinted onto the stellar spectrum to create the simulated total transmission spectrum, $I_\mathrm{tr}$:

\begin{align}
    I_\mathrm{tr}(\lambda , t) = I_\star(\lambda, t) - I_\mathrm{b}(\lambda , t)
\end{align}

where $I_\star$ is the template that represents the out-of-transit or baseline stellar spectrum, and $I_\mathrm{b}$ is the stellar flux that is blocked by the planetary disk and its surrounding atmosphere i.e. the spectrum of exoplanetary absorption features. Spectra are initially in units of energy/wavelength bin (erg cm$^{-2}$ s$^{-1}$ cm$^{-1}$) and are converted into units of photons per pixel at a later step. At the end, planetary and stellar components of $I_\mathrm{tr}$ are Doppler-shifted according to the radial velocities of the planet, star, and barycentric velocity.

To calculate $I_\mathrm{b}$, the planet can be thought of as a single opaque disk with a wavelength-dependent variable for the blocked radius $R_\mathrm{b}(\lambda)$. The area of this radius covers a region of the stellar surface, with a representative spectrum of the blocked stellar region given by $I_{\star \mathrm{b}}(\lambda, t)$ for each exposure during the transit. This blocked spectrum varies in intensity and rest frame from the overall stellar spectrum $I_\star$ due to the limb darkening and rotation of the stellar surface, with the varying amount of occultation during the ingress and egress also accounted for here. Thus, the radiation that is blocked by the planet and its atmosphere $I_b$ can be expressed as:

\begin{align}
    I_\mathrm{b}(\lambda , t) = I_{\star \mathrm{b}}(\lambda, t) \cdot \left(\frac{R_\mathrm{b}(\lambda)}{R_\star}\right)^2\
\end{align}

Our spectral simulator makes the following assumptions: (i) the star has no surface inhomogeneities such as stellar spots or flares and emits a constant limb-darkened spectrum; (ii) the planet surface and atmosphere are homogeneous and produce a time-independent planetary spectrum; (iii) the observation is photon noise limited, and other sources of noise can be ignored. Extinction is accounted for according to the air mass at each exposure. Telluric lines are imprinted in the combined spectrum of star and planet using a synthetic model spectrum of the standard atmosphere at Paranal, which was modeled using Molecfit \citep{smette_molecfit_2015,kausch_molecfit_2015} and then scaled to the airmass value (without accounting for a scaling of H$_2$O lines with the amount of precipitable water). The model also accounts for the Rossiter-McLaughlin effect as the simulation removes the received stellar flux and Doppler shifts this spectrum according to the projected rotational velocity of the stellar surface behind the planet, before subtracting it from the overall spectrum.

All spectra were simulated for observations in the \crires K-band (\SIrange{2.0}{2.5}{\micro\meter}) using wavelength setting \texttt{K2148}. Instrumental effects acting on the spectra include the instrumental profile, which is assumed to be of Gaussian shape with a FWHM corresponding to a spectral resolution of 100,000 (i.e. approximately 3 pixels), and the spectra are scaled according to the blaze functions measured for \crires. For more technical details regarding the specifications of the \crires instrument, please consult the publicly available \crires User Manual from ESO.\footnote{\url{https://www.eso.org/sci/facilities/paranal/instruments/crires/doc.html}}

The expected SNR/pixel for a given observation was estimated using values computed using the adaptive optics (AO) setup with the CRIRES Exposure Time Calculator\footnote{\url{https://etc.eso.org/observing/etc/crires}}, which were fitted to yield the following relation for the K-band: 

\begin{align}
    \mathrm{SNR} &= 247.31\cdot 10^{-m_K/5} \cdot 10^{\kappa(1 - \mathrm{AM})/5} \cdot \sqrt{t_\mathrm{exp}}\,,
\end{align}

with the K-band magnitude $m_K$, the extinction coefficient assumed to be $\kappa = 0.05$, the airmass AM, and the exposure time $t_\mathrm{exp}$ in seconds. After scaling the spectra according to the SNR for the desired exposure time, artificial photon noise drawn from a normal distribution (with a width of the SNR) was added.

Throughout a transit, the planet's atmospheric lines will be smeared over multiple detector pixels due to the planetary motion. We simulate this effect for longer exposures by computing spectra for several shorter subexposures (assuming no additional readout overhead) that are then co-added with the proper Doppler shifts. 

 The generated model data differs from real data in that it does not go through the standard \crires data reduction pipeline. In reality, raw data from \crires is 2D (in the dispersion and spatial directions) and the reduction pipeline converts these observations into 1D spectra. Our simulations generate 1D spectra and so the 2D extraction step is not needed. Several other effects that are handled by the reduction pipeline – such as correction for dark current, bias, cosmic rays, and other instrumental effects – are not included in the model data. 

 At each time resolution, 100 realisations of the same transit event were generated in order to create a larger statistical sample to analyse. Between realisations, only random data noise and start time of the observation relative to the transit mid-point (varying by 1 exposure length) changes. Model observations were also generated for imagined variations of this reference case in order to investigate if and how results would vary for other theoretical planets: one case where the planetary signal was halved (fainter target); one case where the transit duration was halved (same semi-major axis, but shorter period); and one case where the target was observed at a lower spectral resolution ($R$ = 50,000 instead of $R$ = 100,000). Observational data was simulated for each of the four cases at nine different time resolutions, ranging from approximately 2 to 50 minute-long exposures, for 100 simulated realisations of the same transit event; in total, this results in 3,600 simulations (see Section \ref{section:Results}).

\section{Analysis}\label{section:Analysis}

  It is important that the simulated observational data is treated as equally as possible to genuine observational data in order to retain realism. As such, the subsequent analysis for the model data follows standard analysis procedures as closely as possible, starting with the removal of stellar and telluric contamination followed by cross-correlation.

    \subsection{\sysrem}\label{subsection:SYSREM}

  The first step is to isolate the exoplanetary atmosphere spectrum. As described in subsection \ref{subsection:OptimisationProblem}, the planetary component of the transmission spectrum can be distinguished from the stellar and telluric components by its comparatively large radial velocity shift. This shift is sufficiently significant that the movement of planetary lines should be notable from exposure to exposure across the transit (to an extent that depends on the time resolution!) while stellar and telluric lines are systematically present with nearly negligible offset between exposures. For example, in the generated spectra for our reference case, the stellar lines shift by 0.82 km/s during the observation (due to barycentric velocity); the telluric lines are completely stationary; and the planet by $\pm$16.45 km/s throughout the transit. The difference between these can be used to identify and remove non-planetary lines.
 
    The \sysrem algorithm first described in \cite{tamuz_correcting_2005} was originally designed to correct the systematic variations of light curve observations, and it has since been used in a wide range of exoplanet cross-correlation studies for removing stellar and telluric features \citepeg{birkby_detection_2013, birkby_discovery_2017,hawker_evidence_2018,sanchez-lopez_water_2019}. A major advantage of this algorithm is that it does not require any a priori knowledge of the observational features that might influence the measurements. In the visible wavelength regions that are less contaminated by tellurics, alternative approaches to using iterative algorithms like \sysrem generally include methods that model tellurics with radiative transfer codes (such as Molecfit, mentioned in Section \ref{subsection:SimsObs}) and then mask the affected spectral regions before dividing out the stellar signal using baseline measurements from before or after the transit \citepeg{allart_search_2017,mccloat_atmospheric_2021,mounzer_hot_2022}; however, in the infrared wavelengths, the spectrum is so flooded with tellurics that this is often not a practical approach.

\sysrem achieves its goal by representing the data with a model $f$ that consists of two components: a spectrum $S$ that is constant in time (but dependent on wavelength) and the time variation $A$ (constant in wavelength but dependent on time) as expressed by:

\begin{equation}
    f(\lambda, t) = S(\lambda)\cdot A(t)
    \label{eq:sysrem_1}
\end{equation}

Both components $S$ and $A$ are fitted consecutively to the data by minimising the sum of the residuals squared
, meaning the spectrum $S$ will contain both the stellar and the telluric spectrum, and the time variation $A$ includes several factors like the seeing variation, changes in air mass, etc. The planet atmosphere signal is, however, not included in the model as it changes in time due to the Doppler shift. As such, after removing the \sysrem model $f$ from the observations, one should be left with the planet signal, the residuals of time variability not associated with systematic Doppler shift (e.g. variations of seeing), and the noise. This method is closely related to the method of principal component analysis (PCA) and can be interpreted as removing the $N$ largest components from the observations.

Usually this algorithm is iterated several times to gradually remove features present at the same wavelengths. The number of iterations must be carefully selected; an insufficient number of iterations can result in unwanted contamination being retained, yet an excess of iterations will eventually remove the planet signal, particularly in the phases where the Doppler-shifted planetary lines fall on the wavelengths of strong telluric features. Since the algorithm is not formulated as an optimisation problem, there remains an element of subjectivity in selecting the number of iterations. This choice is not always obvious, so one example of a more robust approach involves injecting an artificial (known) signal and testing how many PCA iterations were required to remove it as tested by \citet{cheverall_robustness_2023}. In this work, we tested a wide range of \sysrem iterations in order to track the impact of selecting this iteration number.


    \subsection{Cross-correlation}\label{subsection:CC}

Using cross-correlation is by now a standard practice in the field of characterising exoplanetary atmospheres with high-resolution spectroscopy \citepeg{birkby_detection_2013,kok_identifying_2014,birkby_discovery_2017,brogi_framework_2017,hawker_evidence_2018, giacobbe_five_2021, prinoth_titanium_2022}. After applying \sysrem to remove stellar and telluric signals, one is left with a minute planetary signal buried within the noise in the form of residuals for each exposure. It is these residuals that are then cross-correlated with a template spectrum of the target atmosphere at a range of radial velocity offsets, which effectively combines the signal from all planet atmosphere absorption lines and boosts the SNR of the atmosphere detection. 

The template is a simulated transmission spectrum computed using a highly simplified planet atmosphere model and atomic/molecular data of the species expected to be present at the given physical conditions. Provided that the modelled features in the template do in fact appear in the planetary spectrum, a cross-correlation of the observation and template should confirm their presence even though the predicted relative strength of different lines may be incorrect. For real data, the cross-correlation is carried out between this template and the reduced observational data. In this work, the cross-correlation is carried out between the template and the reduced simulated data, for which the planetary signal is modelled using the very same template (created with \petitradtrans as detailed in subsection \ref{subsection:SimsObs}). This means that in our subsequent analysis, we know that the features definitely exist in our data, giving us the largest possible cross-correlation peak. What we evaluate here is therefore not merely the existence of a cross-correlation detection, but rather its significance relative to the noise of our simulations across the parameter space we intend to explore. 

Generally, cross-correlation studies are performed using templates of individual atomic or molecular species at a time, and the strength of each species/template's detection is calculated respectively. Once this is completed, it can be possible to use this information to infer an overall atmospheric composition using a suite of retrievals based on the detection strength of each chemical species, 
and the fit of this retrieved model can be measured by then cross-correlating the observed spectrum with the retrieved spectrum as shown by e.g. \citet{brogi_retrieving_2019,gibson_detection_2020,lesjak_retrieval_2023,prinoth_time-resolved_2023}. Effectively, in this work, as we are certain that we know the true global transmission spectrum a priori (i.e. our template), it is this final step of likelihood-fitting that is being simulated and whose performance is measured.

The weighted cross-correlation function (CCF) dependent on velocity $v$ and time $t$ can be expressed as:

\begin{align}
    \mathrm{CCF}(v, t) = \sum _{i = 0}^N \frac{R_i(t) \cdot M_i(v)}{\sigma_{R,i}(t)^2}\,
\end{align}
where $i$ is the pixel index, $R$ is the spectrum of residuals (after \sysrem), $M$ is the template spectrum from \petitradtrans, and $\sigma _R$ is the uncertainty of $R$. $\sigma_R$ is usually obtained from the pipeline, but for our model, this is estimated by assuming a constant error of every pixel and subsequently applying the same correction during the steps of normalisation and \sysrem to these as to the data.

\begin{figure}
    \centering
    \includegraphics[width=\columnwidth]{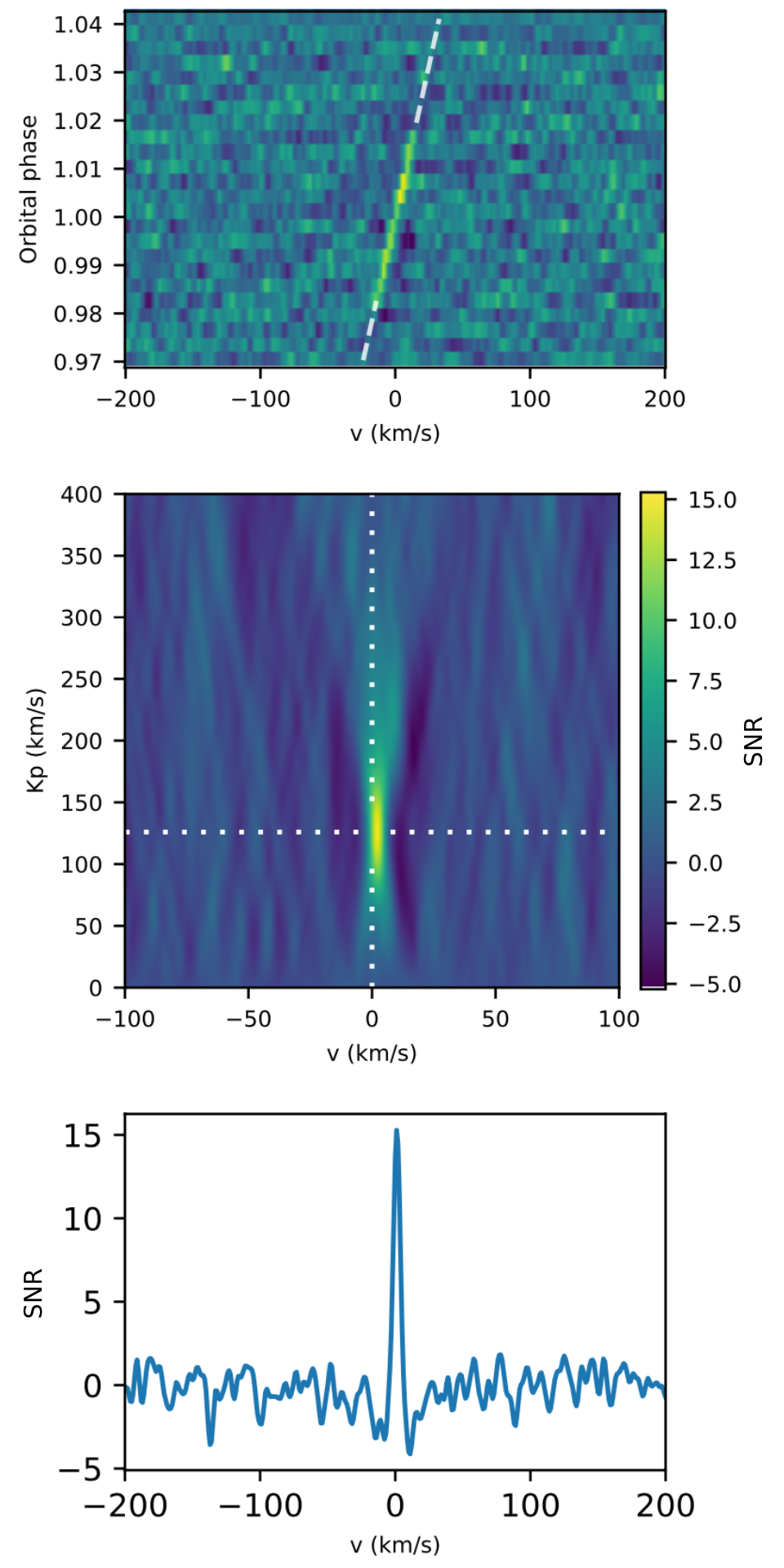}
    \caption{Examples of data plots at three different points in the analysis. (a) Top panel: the exposures are stacked along the vertical axis in time (where orbital phase $\phi=1.0$ is the middle of the primary eclipse). The dotted white lines indicate before and after the transit. (b) Middle panel: the $K_\mathrm{p}-v_{\mathrm{sys}}$ detection map, where all values have been shifted into the planetary rest frame. The white dotted lines indicate rest frame system velocity ($v_{\mathrm{sys}}=0$ km/s) and the semi-amplitude of the radial velocity of the planet ($K_\mathrm{p}=126$ km/s) (c) Bottom panel: an example of what the cross-correlation function looks like at $K_\mathrm{p}=126$ km/s i.e. at the horizontal white line from (b).}
    \label{fig:ccf1}
\end{figure}

 For each exposure, the template spectrum is shifted in \SI{1}{km/s} steps across a range of $\pm$200 km/s around the systemic velocity $v_{\mathrm{sys}}$ of the host star. The top panel of Figure \ref{fig:ccf1} shows a plot of all exposures across the planet's orbital phases along the vertical axis (i.e. time, where each row is one exposure). At each $v_{\mathrm{sys}}$ step along the horizontal axis, the cross-correlation value can be determined with the \sysrem residuals of each exposure, where a high value indicates matching spectral features between those in the observation and those in the template spectrum. Thus, the transit of the planet reveals itself as a diagonal line of high cross-correlation values: the first, bottom rows of exposures show no significant cross-correlation peak (pre-transit). This is followed by a line of high cross-correlation values, which is slanted due to the orbital motion of the planet being first blueshifted (negative velocities) and then redshifted (positive velocities). Finally, the top rows of exposures again show no significant cross-correlation peak (post-transit).
 
By shifting every cross-correlation function in this plot to the planetary rest frame, the slanted line becomes vertical. The shift corresponds to the planetary radial velocity semi-amplitude ($K_\mathrm{p}$ = 126 km/s, which can be calculated from orbital parameters of WASP-127 b) and collapsing the plot vertically will then amplify the cross-correlation peak. Without a priori setting the correct $K_\mathrm{p}$, one can also try different values of $K_\mathrm{p}$ from a large range (e.g. $0-400$ km/s); this gives results that will each be similar to the top panel of Figure \ref{fig:ccf1} but with different slant angles, and realisations producing more vertical lines will result in narrower and higher peak after collapsing the image vertically. 

Stacking the results of different shifts as sorted by $K_\mathrm{p}$ produces an image similar to the middle panel of Figure \ref{fig:ccf1}. This is a `detection map' as the planetary atmosphere detection will show up as a bright, central region that is centred at $v_{\mathrm{sys}}$ = 0 km/s and $K_\mathrm{p}$ = 126 km/s. Overlaying each $K_\mathrm{p}$ value in a single plot in this way gives an intuitive, visual representation of the detection strength as the brightness indicates where cross-correlation values are high, i.e. where there is a peak in the cross-correlated SNR, as seen at the bottom of Figure \ref{fig:ccf1}.

\subsection{Measure of optimisation} 
    \label{subsection:OptimisationTheory}

Considering this work is in the pursuit of `optimal' observations of an exoplanet atmosphere, one must explicitly establish what this potentially subjective term means. Since the current state of the field is largely focused on attempts to identify and characterise chemical species present in exoplanetary atmospheres using comparisons to templates, the definition employed in this paper for what constitutes as an `optimal observation' is: whatever time resolution results in a cross-correlation detection map, i.e. the $K_\mathrm{p}-v_{\mathrm{sys}}$ plot, with the most significant detection. Using this value as a proxy for `optimal' is valid as it is not only the value arguably most relevant to real observational measurements (as it is the value generally used to determine whether a detection has or has not been made), but also because it will vary with both parameters, as it should improve with both increased SNR per exposure (low time resolution) and with reduced smearing (high time resolution). 

This relationship can be understood more explicitly by considering how to calculate planetary SNR. For a close-in exoplanet, assuming negligible telluric interference, the SNR for the planet can be estimated to the first order to depend on:

\begin{equation}\label{SNR_firstorder}
    \mathrm{SNR}_{p} = \bigg( \frac{s_p}{s_\star}\bigg) \; \mathrm{SNR_\star} \sqrt{N_{\mathrm{lines}}}
\end{equation}

where subscripts $\star$ and $p$ refer to the star and planet respectively; $s$ is signal strength; and $N_{\mathrm{lines}}$ is the number of detected (resolvable) lines in the given wavelength range, which accounts for both how many lines are identified and their depths. By detecting multiple spectral lines, the SNR is boosted by the factor of $\sqrt{N_{\mathrm{lines}}}$ since retaining both depth and plurality of spectral lines gives greater confidence in our identification of each unique combination of line patterns as being those of a particular molecule or species \citep{snellen_combining_2015,birkby_exoplanet_2018}. This is ultimately why SNR is both negatively affected by smearing and positively affected by boosted signal strength: smearing spectral lines reduces the SNR of the detection, as this results in fewer resolvable lines, while an increase in planetary signal strength improves the $s_p/s_\star$ ratio.

The measure of detection strength in each detection map, denoted by \Ss, is given by the maximum cross-correlation value resulting from dividing the cross-correlation peak by the standard deviation of the detection map (calculated by excluding the central peak, i.e. everything that is further than 50 km/s to the left or right in the detection map). Physically, the \Ss-value is therefore an estimate of significance of the peak in each plot in comparison to random fluctuations. In order to confirm that this sampling method is robust against the fact that individual samples in the $K_\mathrm{p}-v_{\mathrm{sys}}$ should not be statistically independent, the performance of this method was compared to that of two other sampling methods (first sampling at only one row, namely the row of the signal's $K_p$, and then sampling at every tenth $K_p$ row to avoid small number statistics). These three sampling methods were found to give very similar results with an average standard deviation of $\sigma = 0.078$ from each other (ranging from $0.007 \leq \sigma \leq 0.192$), indicating that the choice of exact sampling method is of negligible impact.

In order to provide more intuitive results to the reader, the measure of time resolution is given in SNR/exposure in this paper. As described, a high time resolution (shorter exposures) gives lower SNR per exposure, and a low time resolution (longer exposures) gives higher SNR per exposure. Considering that different systems require different total exposure times (depending on e.g. transit duration for a given exoplanet), citing a time resolution as an exposure length of $n$ seconds does not automatically evoke an idea of whether this is high or low; instead, we cite total SNR per exposure to facilitate the comparison of relative exposure cadences across different systems. Here, a low total SNR/exposure represents a high time resolution (short exposures) and a high total SNR/exposure represents a low time resolution (long exposures).

\section{Results}
    \label{section:Results}

\begin{figure*}
    \centering
    \includegraphics[width=\textwidth]{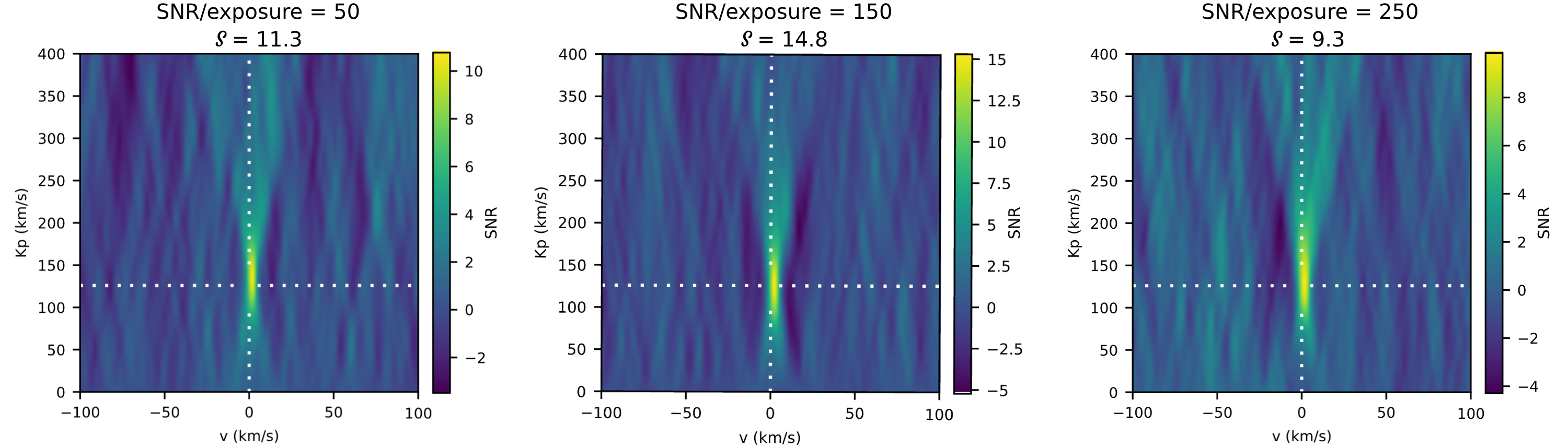}
    \caption{An example of three $K_\mathrm{p}-v_{\mathrm{sys}}$ detection maps for different time resolutions (`high' resolution for SNR/exposure = 50, `medium' for SNR/exposure = 150, and `low' for SNR/exposure = 250) for the WASP-127 b case at \sysrem = 5. For each detection map, its respective measure of detection strength \Sss is cited (see Section \ref{section:Results} for definition) with a higher \Ss-value implying a stronger cross-correlation detection.}
    \label{fig:timeResComparison}
\end{figure*}

For each hypothetical case's `target' (reference study case and its variations), simulations were generated for nine different time resolutions. These ranged from the highest resolution at 186 exposures of 132 seconds each (plus 14 seconds overhead for each exposure) to the lowest resolution at 9 exposures of 2,998 seconds each for the entire observation including baseline out-of-transit exposures. The nine time resolutions resulted in SNR/exposure ranging between 50 and 250, increasing in increments of SNR = 25 for each set of longer exposures. For each combination, we computed 100 realisations of the same transit event, and the results of these realisations were averaged in order to obtain a mean solution as a reference and to assess differences between individual realisations. Each set was then filtered by \sysrem and $K_\mathrm{p}-v_{\mathrm{sys}}$ detection maps were generated after each \sysrem iteration. An example of what different single-realisation detection maps looked like for different time resolutions can be seen in Figure \ref{fig:timeResComparison}. 

 An overview of the different parameters for each case is detailed in Table \ref{tab:ObsParameters}. The four cases were constructed as follows:

 (i) \textit{Reference case}: This refers to the fiducial planet that is representative of the type of target that is nominally observable with our selected instrument. For the target, we use the system and transit parameters of exoplanet WASP-127 b and its host star.
 
 (ii) \textit{Half signal}: This refers to the case of halved planetary signal strength, which is achieved by reducing the line depths of the planetary model by 50\%. This case effectively simulates a range of circumstances that could result in the reduction of signal strength, such as planet size, host star magnitude, distance, data lost to weather, etc.

 (iii) \textit{Half transit duration}: This refers to the case of the halved transit duration, which is modelled by reducing the orbital period only. The transit duration was nominally 4.35 hours, and so for this case, it was reduced to 2.18 hours. For this case, the same number of exposures and the same exposure lengths were used; effectively, the change here is that a smaller number of exposures contains planetary signal.

 (iv) \textit{Half spectral resolution}: This refers to the case of the halved spectral resolution, simulating observations taken at $R$ = 50,000 instead of $R$ = 100,000. All other instrumental effects remain the same. 

\begin{table}\small
    \centering
        \caption{Table of differences between the four hypothetical cases, noting what their respective deviations are from the reference case (which uses parameters from Table \ref{tab:SimsParameters}).}
    \begin{tabular}{ lrr } 
     \toprule
     \textbf{For all cases} & & \\
        \midrule
       Parameter & Symbol & Value \\
     \midrule
    Maximum time resolution (t.r.) & $n_\mathrm{exp}$; $t_\mathrm{exp}$ [s] & 186;132 \\
    Maximum t.r. total observation & $t_\mathrm{obs}$ [hours] &  6.82 \\

    Minimum t.r. &  $n_\mathrm{exp}$; $t_\mathrm{exp}$ [s] & 9; 2,998 \\
    Minimum t.r. total observation & $t_\mathrm{obs}$ [hours] &  7.50 \\
    \midrule
    \midrule
    \textbf{Reference case} & & \\
        \midrule
    Signal strength & $(R_p/R_*)^2$ & 1.70$\times$10$^{-4}$ \\
    Transit duration & $t_\mathrm{t}$ [hours] & 4.3529 \\
    Spectral resolution & $R$ & 100,000 \\
    \midrule
    \midrule
    \textbf{Half signal case} & & \\
        \midrule
    Signal strength & $(R_p/R_*)^2$ & 8.50$\times$10$^{-5}$ \\
    Transit duration & $t_\mathrm{t}$ [hours] & 4.3529 \\
    Spectral resolution & $R$ & 100,000 \\
    \midrule
    \midrule
    \textbf{Half transit duration case} & & \\
        \midrule
    Signal strength & $(R_p/R_*)^2$ & 1.70$\times$10$^{-4}$\\
    Transit duration & $t_\mathrm{t}$ [hours] & 2.1765 \\
    Spectral resolution & $R$ & 100,000 \\
    \midrule
    \midrule
    \textbf{Half spectral resolution case} & & \\
        \midrule
    Signal strength & $(R_p/R_*)^2$ & 1.70$\times$10$^{-4}$ \\
    Transit duration & $t_\mathrm{t}$ [hours] & 4.3529 \\
    Spectral resolution & $R$ & 50,000 \\
     \bottomrule
      & & \\
     
    \end{tabular}

    \label{tab:ObsParameters}
\end{table}

\subsection{Idealised case: no stellar/telluric signal, no SYSREM}\label{subsec:idealcase}

For certain types of observations, removal of stellar and telluric lines is significantly less complicated and does not require algorithms like \sysrem. For example, in the optical regime, telluric lines are much more sparse as the dense bands of molecular absorption do not appear until regions of longer wavelengths. As mentioned, at these spectral ranges, telluric removal can often be done by simply masking out the comparatively small number of telluric lines as needed while the stellar signal can be divided out based on data from out-of-transit exposures. However, in the infrared regime, this is not possible due to the vast breadth of wavelength regions that would have to be masked – and so, having a method for removing stellar and telluric signal in the more targeted way that \sysrem offers is very necessary.

\begin{figure}
    \centering
    \includegraphics[width=\columnwidth]{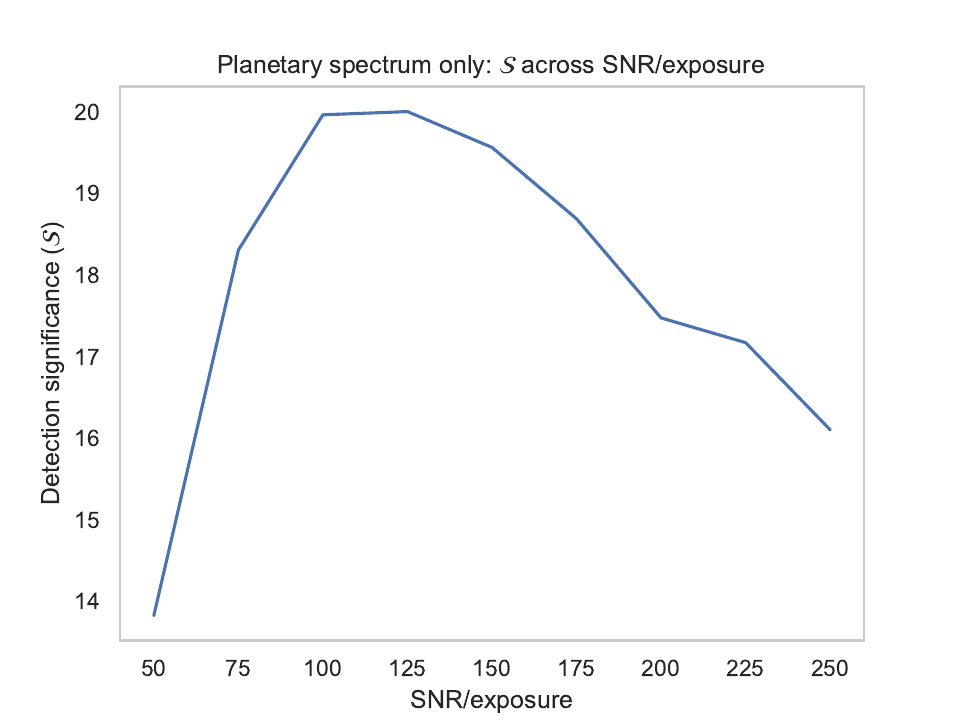}
    \caption{Idealised reference case of no stellar or telluric signal, averaged over 100 realisations. Here, all stellar and telluric contamination are assumed to have been removed successfully (or be otherwise absent, e.g. for telluric contamination in space-based observations) through methods other than \sysrem, which is how stellar and telluric signal will be removed for the subsequent non-ideal cases.}
    \label{fig:nostellurics}
\end{figure}

This means that for any observations that are simulated specifically for a spectrograph like \crires, algorithms like \sysrem make up an important step in analysis that is needed to be retained for realism, but this also means that any and all interpretations of the results have to be made through the lens of this method. In order to disentangle our results from the effects of \sysrem, we first explore an idealised case (based on the same parameters of the so-called reference case) where the stellar spectrum is replaced with a simple limb-darkened black body and where no tellurics have been added, also simulated over 100 realisations. This is essentially a case where one can assume that all stellar and telluric signal have been successfully removed through alternative methods (meaning this case also represents those of space-based observations). Figure \ref{fig:nostellurics} shows a plot for this idealised scenario, demonstrating that under these conditions, the maximum for a trade-off does indeed exist and it lies between SNR/exposure = 100–125.

\subsection{\crires simulations using SYSREM: averaged realisations}\label{subsec:sims-averaged}

\begin{figure*}
    \centering
    \includegraphics[width=\textwidth]{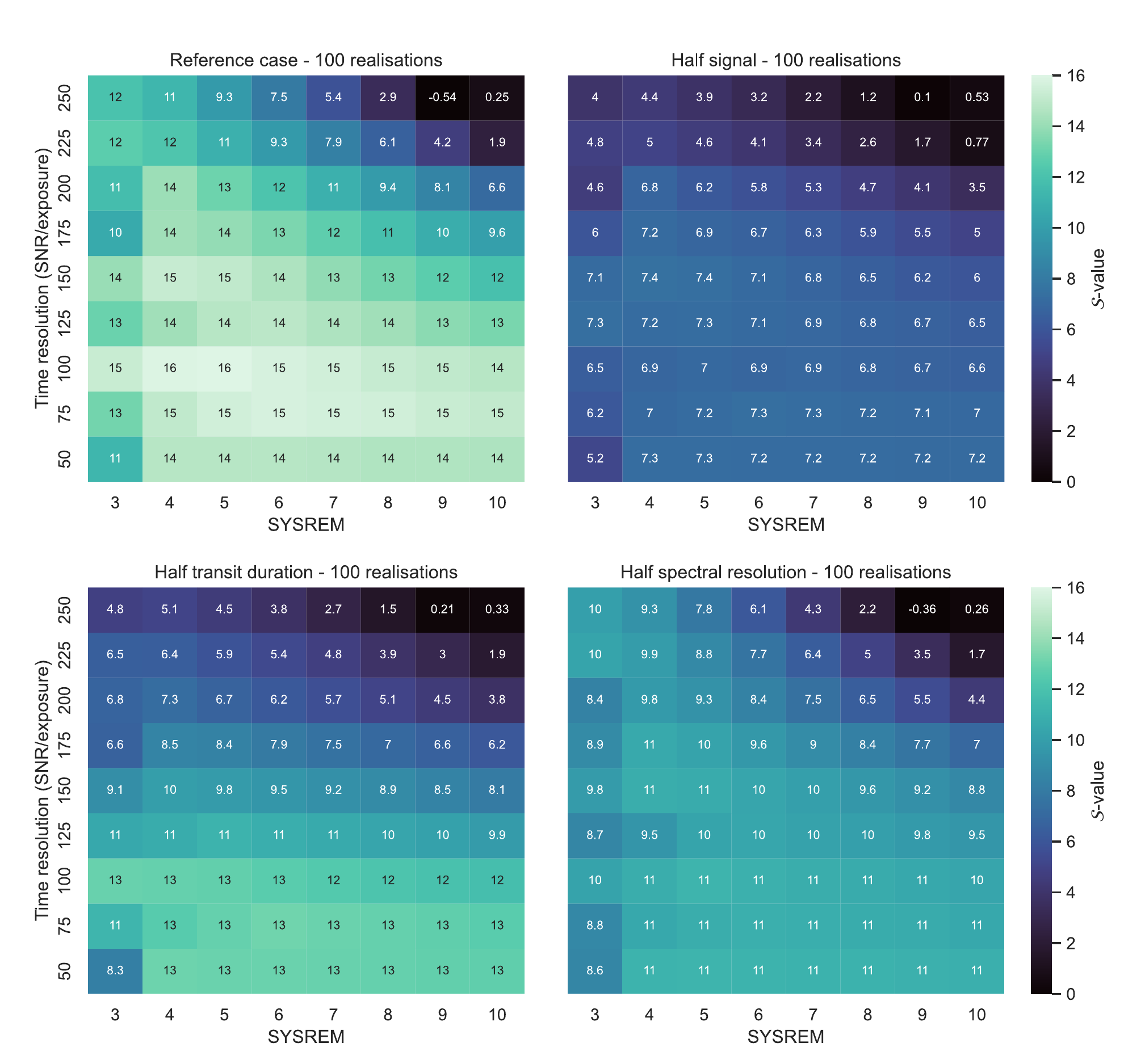}
  \caption{Plots showing how the cross-correlation detection strength, \Ss, varies across time resolution and \sysrem iterations for four hypothetical cases averaged across 100 realisations of the same observing night. (i) Top left: the reference case of a hypothetical observation ($R$ = 100,000) based on fiducial planet WASP-127 b; (ii) top right: a scenario where the reference case's planetary signal strength is halved; (iii) bottom left: a scenario where the reference case's transit duration is halved; (iv) bottom right: a scenario where the reference case is observed with half the spectral resolution ($R$ = 50,000). In these plots, a large number of short exposures corresponds to a lower SNR/exposure (bottom of vertical axis) and a small number of long exposures corresponds to a higher SNR/exposure (top of vertical axis).}
    \label{fig:s-grid_100night}
\end{figure*}

 The value of \Sss for each detection map was recorded, and the average \Ss-values over 100 transit realisations for the four cases were plotted on a grid as heatmaps as seen in Figure \ref{fig:s-grid_100night}. The brighter areas indicate higher values of \Sss for that particular combination of time resolution (SNR/exposure value) along the vertical axis, and the number of \sysrem iterations along the horizontal axis.  As the first two \sysrem iteration are not sufficient to remove telluric features, these were not included and only data from iterations 3–10 were kept for the results seen in this section.

For the reference case (top left), i.e. simulations of planet WASP-127 b over 100 realisations, there is a well-defined parametric space of maximum \Ss-values that indicates robust detection located in the range of approximately SNR = 50–150 across 4–7 \sysrem iterations. Specifically for SNR = 75–100, \Ss-values peak at \Sss $\gtrsim$ 15. Horizontally, \Ss-values appear to peak around \sysrem $\approx$ 4–5 iterations, with subsequent iterations seeing a steady decrease across all time resolutions (as per the `triangular' shape of the heat map). This is an indication that at lower time resolutions (high SNR/exposure), even a low number of \sysrem iterations will start to destroy the planetary signal.

For the case of the halved signal (top right), \Ss-values are approximately halved (\Sss $\gtrsim$ 7) compared to the reference case as one might intuitively expect – but notably, while the numerical values of \Sss have scaled accordingly by approximately half, the location of the parameter space maximum has not moved significantly. In this case, the approximate best range of SNR is also SNR = 50–150 with a similar decay following \sysrem iterations beyond \about 4. In this case, the  maxima has not shifted in location or breadth, but only in strength.

For the case of the halved transit duration (bottom left), the number of exposures that are taken during the transit is less than for the previous two cases; the total observing time and the exposure length vs. exposure time configuration have not changed, but the number of baseline exposures that are taken in out-of-transit has increased since the transit is shorter in time. Here, the region of maximised \Ss-values is shifted down towards higher time resolutions, i.e. larger number of exposures of SNR $\lesssim$ 100, with the highest \Ss-values here being approximately \Sss $\sim$ 13. This can be explained by the interpretation that as the radial velocity of the planet increases, the smearing effect per exposure also increases, and the need for a larger number of exposures becomes more pertinent. As the planet transit time is reduced, so is the total SNR collected across the transit, which is why the maximum \Ss-values of this case is less than that of the reference case. For this case, \sysrem performance still tends to decay at higher iterations, but the `slope' of this triangular shape is flatter than for the previous two cases.

For the case of the halved spectral resolution (bottom right), the overall \Ss-values are globally lower compared to those of the reference case and the halved transit case, with highest values of \Sss $\sim$ 11. These \Ss-values are still inevitably higher than for the case of the halved signal, but the contrast between this case and the reference case greatly emphasises the benefit of high spectral resolution for ground-based observations. While this case also shows \sysrem performance decay with more iterations at longer exposures, it also shows a much more uniform distribution of \Ss-values across the vertical axis. This reflects the reduced impact of the smearing due to the change of planet line-of-sight velocity in comparison to the spectral resolution of the instrument. In this case, the detection significance is already low, and shorter exposures are not able to recover lost spectral resolution.  In a sense, as the contrast of planetary spectral lines is lost due to lower resolution, this loss dominates the detection strength of an observation more significantly than any strength lost or gained due to changes in time resolution. As such, the loss of sensitivity to atmospheric features due to reduced spectral resolution cannot be fully recovered with improved time resolution.


\begin{figure*}
    \centering
    \includegraphics[width=\textwidth]{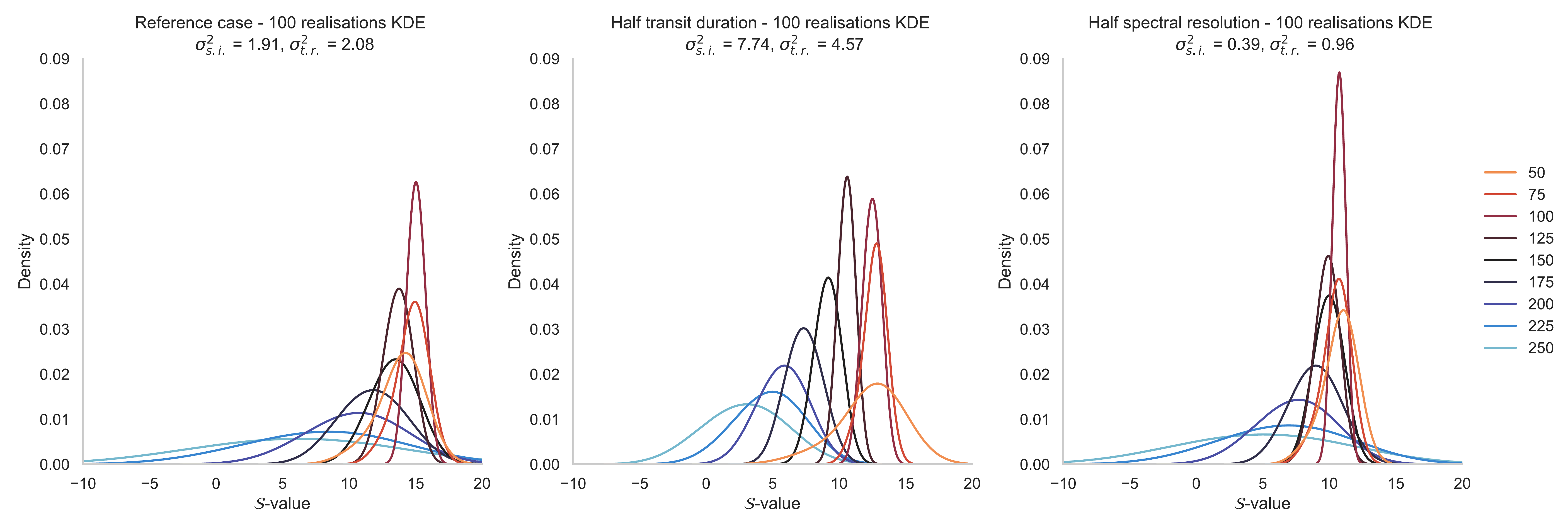}
    \caption{Kernel density estimates (KDE) of \Ss-values for all different time resolutions SNR/exposure = 50–250 across all SYSREM iterations, seen for the the reference case (left) compared to the halved transit duration (middle) case and the halved spectral resolution range (right) using the data presented in Figure \ref{fig:s-grid_100night}. The variances are calculated for each case across all time resolutions at a selected iteration of SYSREM = 4 ($\sigma^2_{s.i.}$) and across all SYSREM iterations for time resolutions SNR/exposure = 50-175 ($\sigma^2_{t.r.}$).}   \label{fig:KDE_100nights_3cases}
\end{figure*}

The effect of smearing being more pronounced for faster targets with shorter transits yet less pronounced for observations with lower spectral resolution can also be seen by plotting kernel density estimates (KDE) for our data sets as in Figure \ref{fig:KDE_100nights_3cases}. In this plot, the range of \Ss-values is shown for the halved transit duration case and for the halved spectral resolution case compared to the reference case. 

Here, the reference case's time resolutions of SNR/exposure = 75, 100 is shown to indeed return the highest \Ss-values, with SNR/exposure = 50 faring next best, followed by SNR/exposure = 125, 150. However, for the halved transit duration case, the range of \Ss-values are much more spread; here, SNR/exposures = 50–100 all fare similarly well (as confirmed by the heatmaps) but subsequent time resolutions quickly deteriorate with each successive time resolution resulting in a decreased peak of \Ss-values due to increased smearing. Comparatively, for the halved spectral resolution, this is not true; here, SNR/exposure = 50–100 still results in the highest \Ss-values but SNR/exposure = 125–175 are not far behind. For all three cases, the time resolutions of SNR/exposure = 200–250 fares significantly worse as is also seen in the heatmaps of Figure \ref{fig:s-grid_100night}, in part due to the diminished success of SYSREM at higher iterations.

This spread can also be identified by calculating the variance ($\sigma^2$) of these data sets. A pronounced smearing effect should register as a larger variance across the data set (due to larger differences between performance of shorter versus longer exposures) and a diminished smearing effect should register as a smaller variance. Two values for variance are given in the titles of the plots of Figure \ref{fig:KDE_100nights_3cases}. The first value $\sigma^2_{s.i.}$ comes from calculating the variance across all time resolutions in a single SYSREM iteration (effectively the variance of \Ss-values in a single column of the heatmaps of Figure \ref{fig:s-grid_100night}). Here, $\sigma^2_{s.i.}$ is calculated for SYSREM = 4 as this iteration consistently performs well for the three respective cases. The second value $\sigma^2_{t.r.}$ is a calculation of variance across time resolutions for all SYSREM iterations (effectively a measure of how much the KDE plots for different time resolutions shift). This value represents the amount of variation between each row of the heatmaps – and considering all SNR/exposure $\geq$ 200 fare consistently badly due to poor SYSREM performance at higher iterations, $\sigma^2_{t.r.}$ excludes SNR/exposure = 200–250. For both $\sigma^2_{s.i.}$ and $\sigma^2_{t.r.}$, the trend seen in the heatmaps holds: the smearing effect is larger (higher $\sigma$) for the halved transit case, and smaller (lower $\sigma$) for the halved resolution case.

\subsection{\crires simulations using SYSREM: comparison between single realisations}\label{subsec:sims-single}

\begin{figure*}
    \centering
    \includegraphics[width=0.95\textwidth]{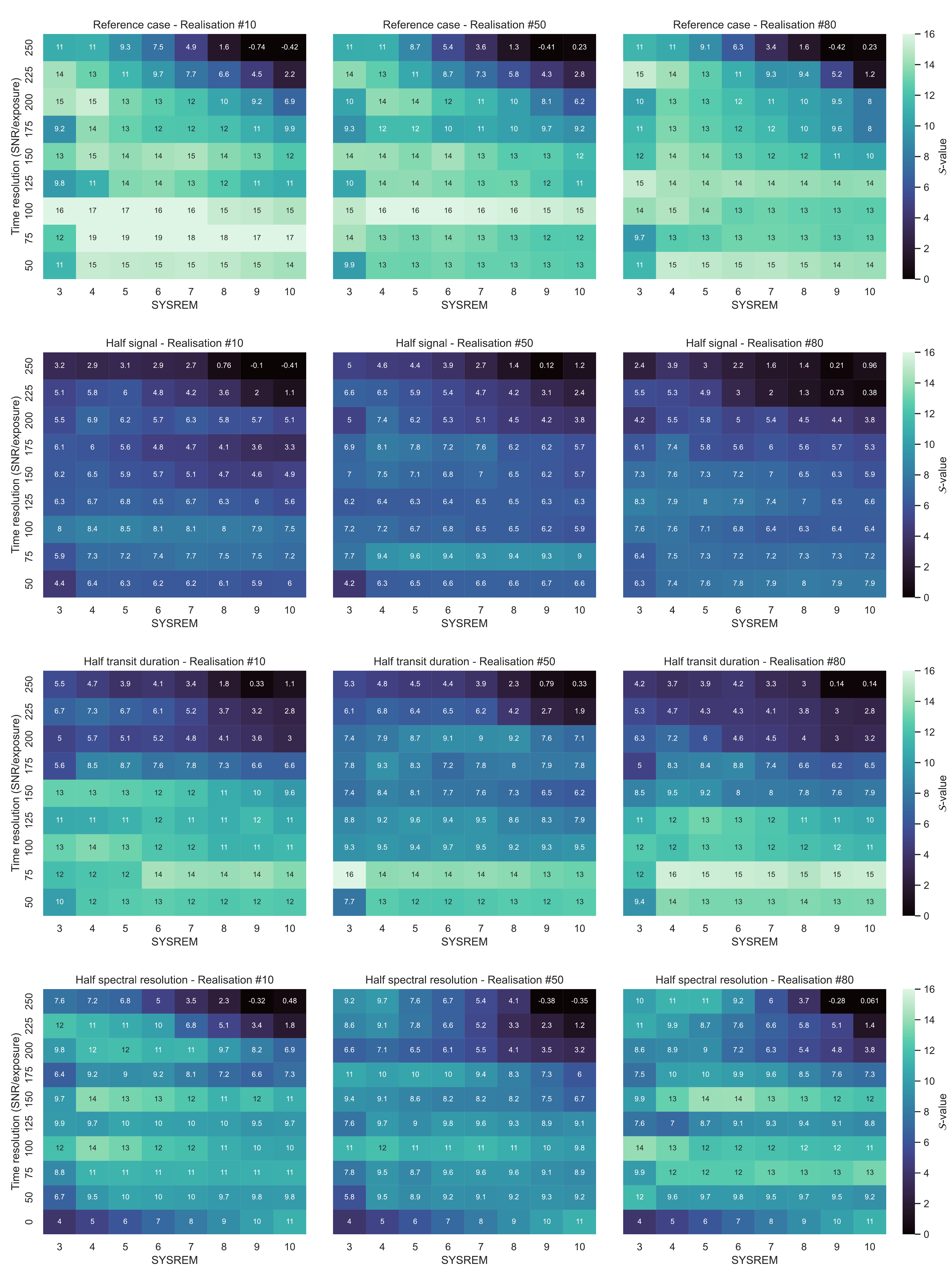}
    \caption{Plots showing how the cross-correlation detection strength, \Ss, varies across time resolution and \sysrem iterations for individual realisations of the four hypothetical cases. Each row represents (i) the reference case, (ii) the case of the halved planetary signal, (iii) the case of the halved transit duration, and (iv) the case of the halved spectral resolution, following the same format and parameters as in Figure \ref{fig:s-grid_100night}. Note that some values are beyond the colour bar's scale, which has been locked to facilitate comparison to Figure \ref{fig:s-grid_100night}.}
    \label{fig:s-grid_singlenight}
\end{figure*}

In contrast to these \Ss-values based on detection maps averaged over 100 realisations,  the \Ss-values of a single realisation can be seen in Figure \ref{fig:s-grid_singlenight}. Here, grids of three randomly
selected realisations out of 100 (realisation number 10, 50, and 80) for each case are shown in order to illustrate how much individual realisations can deviate from the mean. 

Importantly, we find that while the individual realisations' maxima are still confined within the range of maxima seen for the averaged realisations, the exact range of optimal time resolutions do not show up as a clear maxima in a single realisation in the same way. This contrast between the individual realisations and the averaged realisations clearly underlines how small the planetary signal is relative to noise and fluctuations associated with observations, meaning that observations from a given night can – and often will – vary significantly from the mean. This can also be seen in the the presence of horizontal stripes of higher \Ss-values in the single realisation plots: as noise was generated anew for each time resolution (vertical axis), each row has its own maxima of peak \Ss-values in a \sysrem range. There are notable variations between each row as each have their own randomly generated noise, impacting each data set independently.

This is an important finding as we can see that two instances of the same exact circumstances – an observation of the same planetary system, with the same instrument, at the same time resolution – can seemingly in the worst case result in either a detection or a non-detection, with favourable or unfavourable random variation governing the outcome. For example, the individual realisations of the reference case (first row of plots in Figure \ref{fig:s-grid_singlenight}) gives a high \Ss-value detection at time resolution SNR/exposure = 75 at \sysrem = 4–5 (which the averaged realisation confirms is a good choice of parameters) for realisation no. 10, but a detection that is comparatively much more marginal for realisation no. 50 and no. 80. Similar variations can be seen between all cases' different individual realisations, indicating that this is a persistent, systematic issue. In order to eliminate the possibility that this result could be merely a manifestation of \Sss being calculated from the peak value (which is itself noisy), this was also tested by using a Gaussian fitting technique, but this method still replicated these results for both the averaged realisations in Figure \ref{fig:s-grid_100night} and the variation seen in the single realisations of Figure \ref{fig:s-grid_singlenight}.

\section{Discussion}
    \label{section:Discussion}

In this paper, we have used observations simulated for a specific instrument and targets. In order to explore how our findings would vary for other cases and conditions, observers may use the same method to simulate spectra for their own specific target, transit event, or instrument to experiment with resulting cross-correlation \Ss-values as presented here. While probing every category of such variations is certainly outside the scope of this paper, there are certain trends that we expect to hold for any high-spectral resolution transit spectroscopy observation, with certain parameters impacting the range of optimal time resolutions. These are discussed below in the context of our results.

\subsection{Summary of the results}

\begin{enumerate}
\item The \sysrem $+$ cross-correlation approach tends to work well with relatively low SNR observations. When planning observations, one should favour higher time resolution versus higher SNR/exposure as long as the data reaches SNR/exposure of 50 or so. While higher SNR per exposure helps mitigate the large variation of detection significance that we see across the individual realisations, we usually do not have the luxury of combining many transits – as such, striving for higher SNR should not happen at the cost of time resolution beyond this point. Another aspect that could motivate longer (higher SNR) exposures is the overhead time that can be comparable to exposure times for large CCD detectors (see subsection \ref{subsec:instruments_overheads}).


\item Chances of detecting planetary signal are increased by using the highest spectral resolution available. The reason for this is two-fold: firstly, higher resolution improves contrast and effectively boosts $N_{\mathrm{lines}}$ (from Equation \ref{SNR_firstorder}). Secondly, high spectral resolution minimises the impact of telluric contamination (for ground-based observations) and opens the possibility of combining multiple transits in a single analysis if one can observe between telluric lines across multiple nights. The downside of high resolution is the difficulty of achieving high SNR and so the compromise must be found for each target and telescope/instrument combination. In our case, the width of the instrumental profile (1 km/s) is a factor of 10 smaller than the variation of the planet line-of-sight velocity during the transit, which increases the potential of reliably detecting atmospheric signal without losing too much light on the slit.

\item The use of \sysrem has a clear impact on results, with variation in optimal time resolution seen both between cases that do and do not use it (the ideal case of subsection \ref{subsec:idealcase} versus the reference case of subsection \ref{subsec:sims-averaged}), as well as between different choices of iterations within the same time resolution and case (along the horizontal axes of Figure \ref{fig:s-grid_100night} and \ref{fig:s-grid_singlenight}). In the results of subsection \ref{subsec:sims-averaged}, \sysrem cleans the spectra of telluric and stellar features sufficiently already after 4–6 iterations. Generally, it can therefore be said that a large number of iterations is seemingly detrimental to the recovery of the atmospheric signal (as SYSREM may then be struggling to identify what is linear and not with such a small number of exposures) and should be avoided. In our cases, \sysrem performance is more affected by the time resolution of the data than by its SNR, leading to point 1 above.
\end{enumerate}

\subsection{Stacking multiple nights}

If one can observe and combine data from multiple transits, this would improve the total SNR for a given target by stacking observations from different nights – in particular if done with the same instrumental setup. Such observations have been done for both transmission and emission studies in the past using a range of ground-based instruments \citepeg{giacobbe_five_2021,kesseli_atomic_2022,scandariato_pepsi_2023}. It has also been shown that stacking multiple observations will be required for future characterisation of Earth-like and super-Earth exoplanets from space \citepeg{kaltenegger_transits_2009,rauer_potential_2011,wunderlich_detectability_2019}, and this has indeed been achieved with recent JWST data \citep{lustig-yaeger_jwst_2023}. However, stacking of ground-based observations in the mid-infrared will require more advanced telluric removal algorithms than \sysrem considering that the barycentric velocity change between nights and weather variations will results in large changes of telluric features between transits.

As shown in Section \ref{section:Results}, there is a notable difference in the distribution of \Ss-values for a single realisation compared to the distribution of the multi-realisation \Ss-values; however, these effectively show an average of multiple observations, which is not the same as stacking observations to amplify the SNR.
Differences in observing conditions (seeing, telluric line strength) may set a limit to the number of transits
that can be properly combined with a gain to the planetary signal detection.

If the quality of observations of individual transits is similar, the stacked data set will be effectively similar to `oversampling' planet transit with more spectra of the same time and spectral resolution. In any case, stacking multiple transits should result in higher values of \Sss provided that \sysrem is modified to be used with such data set.

\subsection{Variable exposure lengths}

In this work, we have explored different time resolutions with simulations ranging from several shorter exposures to fewer longer exposures. However, it may be possible that a variation exists in what should be prioritised during different sections of the transit event; is the trade-off balance the same during ingress/egress as it is during the middle of the transit? If not, further optimisation could be possible if exposure lengths were varied over time, with either shorter exposures during ingress/egress and longer exposures during mid-transit, or vice versa.

Considering that the background spectrum of the star changes fastest in time close to the limb, this might suggest that shorter exposures during ingress and egress would be beneficial. On the other hand, the specific intensity is much lower across the limbs than in disk centre, which may suggest the opposite. This exact balance – and how it may be affected by further parameters e.g. planetary orbit eccentricity \citep{van_eylen_orbital_2019} – could be explored in future work, but because the two effects are working against each other, one could hope that equal exposure times are not far from the optimal strategy. Furthermore, it is reasonable to assume that the effect will nonetheless be significantly smaller compared to other, more dominant factors such as the number of \sysrem iterations. This, alongside the balance of the two opposite influences, discouraged further exploration of variable exposure lengths as we do not expect it to have a significant effect on the detectability of planetary signal for our particular cases.
\subsection{Exoplanet atmospheric features}

Several exoplanet atmospheric features have been found to affect the transmission spectra of exoplanets. Thanks to the progress in the field, we have been increasingly obliged to acknowledge that exoplanetary atmospheres are complicated and three-dimensional systems that should not be overly simplified; more chemically complicated species like aerosols are now known to be commonly found in many types of exoplanetary atmospheres, creating clouds and hazes that make characterisation difficult at the lower atmosphere layers \citepeg{helling_exoplanet_2019,gao_aerosols_2021}, and the formation of other known meteorological phenomena such as precipitation and winds are also being studied in great detail \citepeg{heng_atmospheric_2015,seidel_wind_2020,fortney_hot_2021,loftus_physics_2021}. Recent years have also seen an rise in adopting 3D general circulation models, or global climate models (GCM), for use with exoplanets in order to demonstrate the importance of atmospheric dynamics and stratification for interpretation of exoplanets' transmission spectra \citepeg{ding_new_2019,wolf_importance_2019,way_venusian_2020,fauchez_trappist-1_2022}. 

The cross-correlation detection map is indeed sensitive to many such atmospheric effects, with evidence of high-altitude winds manifesting as broadening or off-sets. The simulations in this work did not take these effects into account, including e.g. rotational broadening – an effect that would be present most notably on tidally-locked planets with shorter orbital periods – which would presumably lead to an increase in a type of smearing that arises separately from choice of time resolution. As such, in the case of strong atmospheric features, there may be a supplementary sensitivity in what time resolutions result in the optimal cross-correlation detection but the exact nature of this relationship would need to be explored before further conclusions can be reached. This relationship may be investigated by testing the dependence of inferred input atmospheric parameters on simulated observing choices as per e.g. Savel et al. (in prep.).

\subsection{Host star activity}

As mentioned in Section \ref{section:Introduction}, cooler stars such as M-dwarfs are likely hosts of exoplanets and thus commonly observed. M-dwarfs are known to be active stars with magnetic fields that may affect close-in orbiting exoplanets \citepeg{morin_large-scale_2010,airapetian_how_2017,kochukhov_magnetic_2021}, and this high level of activity means that they are likely to have starspots appear on their photospheres. While such activity of M-dwarf host stars can in cases be exploited to aid exoplanet atmosphere studies \citepeg{diamond-lowe_ground-based_2023}, there has been concern regarding what the influence may be of starspots present on the surface of host stars during transits – and unfortunately, the fear of such spots creating false spectral features appears to be well-founded especially at lower spectral resolutions \citep{rackham_transit_2018,apai_understanding_2018,barclay_stellar_2021,moran_high_2023,libby-roberts_-depth_2023}. Considering the simulated transmission spectra in this work assumes a uniform stellar disk with no features, the risk of contamination due to potential activity of host stars should be taken into account when considering the validity of such an assumption. The situation could be different for stars of later spectral type, where the impact of the combination of larger level of activity and smaller convective cells must be further explored.

\subsection{Differences between instruments}\label{subsec:instruments_overheads}

By now, there are tens of astronomy instruments available globally of world-class quality that are regularly used to take spectroscopic observations of exoplanets. As discussed in Section \ref{section:Introduction}, the different traits of the instruments across this wide range all provide advantages and disadvantages regarding high or low spectral resolution, ground- or space-based, mirror size, wavelength coverage, and many more. All of these will combine uniquely to create an exclusive case for each given instrument and target, and so it is possible that the range of optimal time resolutions varies distinctly when using different instruments/telescopes. Furthermore, differences between instruments could be taken advantage of in the case that observers explore the possibility of stacking multiple observations from multiple instruments, as done by e.g. \citet{ridden-harper_high-resolution_2023}.

One important aspect to consider when selecting between different instruments is how large the observing overheads are. \crires has the benefit of being able to observe with relatively small overheads, requiring only a few seconds between sequential exposures. Other instruments have notably longer overheads of e.g. 100 seconds for each pair of frames for the Gemini-N/MAROON-X instrument, and either 45 seconds (fast readout mode) or 68 seconds (slow readout mode) for the VLT/ESPRESSO instrument. 

Accounting for overhead time per exposure ($t_{\mathrm{OH}}$) is important because large overheads will eat into valuable light collection during the transit, with a larger number of exposures ($n_{\mathrm{exp}}$) consuming more observing time. The amount of observing time lost to overheads for a given instrument is given by $t_{\mathrm{OH}}(n_{\mathrm{exp}}-1)$, which means that there will be a larger difference between how much total light collection is lost for a high $n_{\mathrm{exp}}$ and low $n_{\mathrm{exp}}$ if the instrument has a high $t_{\mathrm{OH}}$. As such, for instruments with notably large overheads, faint targets requiring a low number of $n_{\mathrm{exp}}$ need to be reasonably slow, and fast targets requiring a high number of $n_{\mathrm{exp}}$ need to be reasonably bright. For targets that have both short transit times and low signal strengths, observers will benefit from using instruments with relatively small overheads instead.

\subsection{Alternatives to PCA-based approaches}

The major drawback of ground-based observations in the infrared is telluric contamination. Its strength is, ironically, the reason why we observe at these wavelengths as the species present in our atmosphere are often those we search for in our targets. Considering that various techniques of removing telluric features such as \sysrem and other functionally-similar tools have several limitations, their iterative nature makes it difficult to control their impact on the planetary atmosphere features. As such, it is very difficult to assess this impact even in the analysis of the cross-correlation results.

An alternative way to approach this problem of isolating the planetary contribution is to instead model the observations as a whole, treating the components (stellar, telluric, and planetary spectra) as unknown functions. This will lead to an inverse problem that can be solved for a global minimum assuming that the orbital elements and the corresponding Doppler shifts of the components are well known for all phases of the transit. While both algorithmically and computationally more complex, an inverse problem approach has the benefit of producing a single global planet transmission spectrum, rather than a set of residuals for each phase as PCA-based approaches do. Such a technique is currently under development by Piskunov et al. (in prep.), building on preliminary theoretical work by \cite{aronson_using_2015}.

\subsection{Implications for previous studies}

With the finding that the detection strength of individual realisations vary significantly from the mean (and from each other), there is a need to assess what this means for current methodologies as the choice of time resolution or number of \sysrem iteration could be the difference between a detection and non-detection.

Many previous studies have acknowledged the lack of consistent treatment and the challenges associated with using iterative detrending methods such as \sysrem, including the potential for false positives  \citepeg{hawker_evidence_2018,cabot_robustness_2019,cheverall_robustness_2023}. Our findings further underline this need for caution, and for all results (new or revisited), care should be taken to avoid the risk of both false positives and false negatives due to noise variation per observing night. One potential mitigation would be to carry out PCA-analysis injection tests (with their proven success record) over multiple realisations to obtain an averaged result similar to our work as this might improve iteration selection – especially for PCA-based approaches that do not use the same noise model for the injection and for the data – but further work is needed to develop and test the performance of such a method.

\subsection{An `exposure triangle' rule for planning observations}

Our results indicate that the concept of a trade-off in the context of transit spectroscopy observations is a multi-dimensional problem. By varying different parameters, the maxima of the `optimal' time resolution's range can in response vary in location, strength, and size – and only clearly so over a large number of realisations. With so many variables that can influence the result, it is not surprising that changes do not manifest in a linear fashion across the breadth of parameters tested.

It is therefore our finding that while observers will fare best by creating their own simulations and investigating specific scenarios as needed, as a rule of thumb, a strategy akin to the so-called `exposure triangle' may be recommended. In traditional photography, the rule of the exposure triangle is approximately that three elements must be considered prior to taking a photograph: aperture (how widely the camera lens' diaphragm opens), shutter speed (how long the camera shutter opens for), and film speed (how much light can be registered on the camera's film or digital sensor). These three controls balance to determine how much light enters (through aperture and shutter speed) based on how sensitive the instrument is (film speed). Depending on several conditions – e.g. the subject's motion, or the lighting of the subject's location – photographers may be restrained in one or more of these dimensions, and considering the exposure triangle allows them to determine how to compensate for those restraints through adjusting one or both of the others. For example, if one is taking a photograph of a fast-moving car, one needs to reduce the shutter speed to capture it without blurring, but can compensate for that loss of light by increasing the aperture or increasing film speed.

By considering an analogous scenario for capturing transiting exoplanets, a similar rule may be employed: if one is restrained by a particular observational parameter – a short transit, a faint target, or an instrument of low spectral resolution – one may consider the other variables to determine approximately which parameters to prioritise or adjust. As such, there is no single range of time resolutions that will consistently deliver optimal results; like photography, the way to centre your exposure within this set of parameters will depend on several factors as discussed in this section and above. As per our results, certain adjustments are fairly intuitive – for example, that smearing effects are more pronounced for shorter periods but less pronounced at lower spectral resolutions – but others require more consideration or sacrifice.

While exact priorities must therefore be considered for each given case, one can still produce a general recommendation for a single target: smearing is more of a concern when the transit duration is short (because the effect is more pronounced) and less of a concern when resolving power is low (because spectral resolution cannot be regained by shorter exposures). Therefore, observers should first consider the parameters of their target system and chosen spectrograph/instrument. Smearing will be of minimal concern when their target has a relatively long transit duration, and when their instrument has a relatively low spectral resolution; in this case, observers should prioritise a longer exposure for maximum SNR. Inversely, smearing will be of maximal concern when their target has a relatively short transit duration, and their instrument has a relatively high spectral resolution; in this case, observers should prioritise a shorter exposure for minimal smearing.

\begin{figure}
    \centering
    \includegraphics[width=\columnwidth]{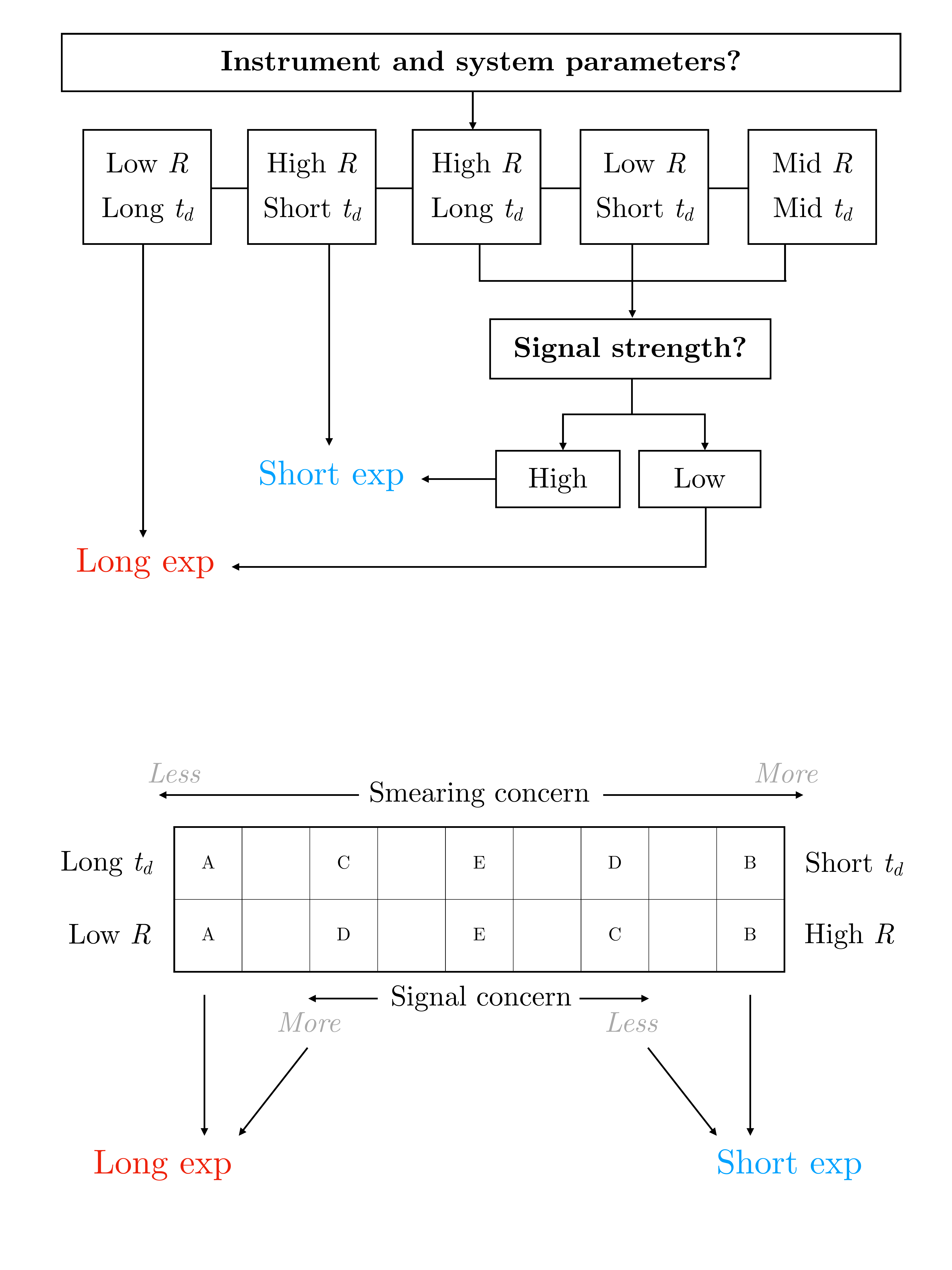}
    \caption{\textit{Top schematic:} A generalised recommendation for a single target. First, observers should consider their instrument and system parameters. For lower spectral resolutions ($R$) and longer transit durations ($t_d$), longer exposures may be prioritised; for the inverse case, shorter exposures are more suitable. For other cases, the choice is determined by what signal strength is possible. \textit{Bottom schematic:} The logic of the top schematic is illustrated here. Smearing is more of a concern for low $R$ and long $t_d$ (case A ) and less of a concern for the inverse (case B). For all other cases (C, D, E), a larger concern for signal (i.e. lower signal strength) encourages longer exposures and a lesser concern encourages shorter exposures.}
    \label{fig:flowchart_triangle}
\end{figure}

In the case where neither of these factors are of notable effect, i.e. either because both factors are effectively balanced (long duration with high $R$, or short duration with low $R$) or because both are of roughly equal detriment (neither factor stands out as being of particularly great impact), one should instead consider the signal strength. Per our simulations, as long as the SNR/exposure can reach $\sim$ 50–75, longer exposures do not appear to improve detections sufficiently to be worthwhile, especially considering that SYSREM clearly performs more poorly with a lower number of spectra (longer exposures). This means that for systems that can guarantee higher signal strength, where observers can trust that they can receive a relatively high SNR even in a shorter exposure due to e.g. high contrast, they may prioritise a shorter exposure in order to retain as much spectral resolution as possible. However, if the signal strength is expected to be weak, and observers know they can only obtain sufficiently high signal strengths by longer exposures, they are recommended to do so in spite of increased smearing.

This generalised recommendation is outlined schematically in Figure \ref{fig:flowchart_triangle}, whose first plot shows a simple flow chart that may be consulted for broader decision-making, with the second plot illustrating the logic of the flow chart structure. What exact values will qualify as being a short/long transit duration, or of high/low spectral resolution, or a high/low signal strength cannot be established explicitly here for every combination of possible targets and instruments; as such, decisions based on this recommendation require a judgement call that must be made by astronomers based on their science goals (e.g. detection or characterisation) and on their global understanding of their upcoming observation. It is this evaluation, where one judges what may or may not be of greatest priority for a particular case, that can be considered as the use of an `exposure triangle' approach in this context. Again, if resources permit, more specific boundaries on time resolution may be calculated if observers choose to create their own simulations following the methodology outlined in this work, but the recommendation above can still be employed as a short-hand guide.

\section{Conclusions}
    \label{section:Conclusions}

In this work, we simulated multiple \crires observations of four hypothetical targets based on fiducial planet WASP-127 b and used the PCA-like \sysrem algorithm – a standard analysis tool in exoplanetary atmosphere studies – to investigate the trade-off between SNR and time cadence when conducting transmission spectroscopy observations. 

(1) We have demonstrated that time resolution significantly impacts the strength of cross-correlation detection for transiting exoplanets. Across our range of data sets, there is a clear trade-off between observing a transit with a small number of higher SNR exposures and a large number of lower SNR exposures, with the range of optimal time resolutions and overall performance varying based on characteristics of the observed system. The exact location, strength, and size of this range varies based on multiple variables, which means that observers need to consider all context at once when evaluating best strategies (akin to the `exposure triangle' rule of traditional photography) and supplementary techniques such as stacking may return different maxima ranges. We caution those planning observations to robustly simulate target systems ahead of time in order to take full advantage of the time limitations imposed both by transit duration and by telescope access.

(2) Based on our simulations, variation in target signal strength do not appear to shift where the location of this maxima exists on the parameter space, but the benefits of prioritising a larger number of exposures should be taken into account for targets with shorter transit times. At lower spectral resolution, the smearing effect associated with faster transiting targets appears to be less pronounced, although a higher spectral resolution clearly results in more significant detection at the optimal time resolution. For the four hypothetical cases explored in this paper, as a rule of thumb, time resolutions of SNR/exposure $\approx$ 50–100 appear to provide relatively high detection strength for all, with no case seeing an optimal resolution of SNR/exposure $>$ 200.

(3) There also appears to be a minimum threshold of time resolution below which \sysrem may destroy the planetary signal after only a small number of iterations, but this exact threshold needs to be determined for specific targets and instrument configurations. For our cases, time resolutions of SNR/exposure $\gtrsim$ 200 appear to be especially sensitive, and SYSREM $\approx$ 4-6 generally seems to yield best results.

(4) We have also found that night-specific variations such as seeing and data noise have a significant impact on the results of that night. As this work has explored ideal cases, this impact is reasonably assumed to be even greater for real-life circumstances e.g. variations in weather conditions. Considering many studies are based on single-night observations, this needs to be accounted for during analysis (including the steps pertaining to stellar and/or telluric removal) and in evaluating the validity of single-night detections.

As the field of exoplanetary atmosphere characterisation makes solid progress in every direction – improved instrumentation quality, analytical methods, data reduction, and modelling methods for generating template spectra for cross-correlation – we can reasonably expect good progress in the relatively near future concerning characterisation of lower-mass exoplanets to match our achievements of higher-mass exoplanets. With such steady developments making rapid strides, it is important to monitor our tools over time to ensure that they do not become obsolete; and so, by monitoring the variation of optimal time resolution for different observations, we can verify that we are continuously obtaining the best possible data for our scientific progress. 

\begin{acknowledgements}
    L.B.-Ch., A.D.R., and N.P. acknowledges support by the Knut and Alice Wallenberg Foundation (grant 2018.0192). O.K. acknowledges support by the Swedish Research Council (grant agreement no. 2019-03548), the Swedish National Space Agency, and the Royal Swedish Academy of Sciences. This research has made use of the NASA Exoplanet Archive, which is operated by the California Institute of Technology, under contract with the National Aeronautics and Space Administration under the Exoplanet Exploration Program. This work has also made use of the following Python packages: \texttt{Astropy} \citep{astropy_collaboration_astropy_2013}, \texttt{iPython} \citep{perez_ipython_2007}, \texttt{Matplotlib} \citep{hunter_matplotlib_2007}, \texttt{NumPy} \citep{harris_array_2020}, \texttt{Pandas} \citep{mckinney_data_2010,team_pandas-devpandas_2023}, \texttt{PyAstronomy} \citep{czesla_pya_2019}, \texttt{SciPy} \citep{virtanen_scipy_2020}, and \texttt{Seaborn} \citep{waskom_mwaskomseaborn_2017}. We also thank the anonymous referee for their comments as their suggestions have contributed significantly to this work.
\end{acknowledgements}
\bibliography{SNRpaper.bib} 

\begin{thebibliography}{127}
\expandafter\ifx\csname natexlab\endcsname\relax\def\natexlab#1{#1}\fi

\bibitem[{Addison {et~al.}(2019)Addison, Wright, Wittenmyer, Horner, Mengel,
  Johns, Marti, Nicholson, Soutter, Bowler, Crossfield, Kane, Kielkopf,
  Plavchan, Tinney, Zhang, Clark, Clerte, Eastman, Swift, Bottom, Muirhead,
  McCrady, Herzig, Hogstrom, Wilson, Sliski, Johnson, Wright, Johnson, Blake,
  Riddle, Lin, Cornachione, Bedding, Stello, Huber, Marsden, \&
  Carter}]{addison_minerva-australis_2019}
Addison, B., Wright, D.~J., Wittenmyer, R.~A., {et~al.} 2019, Publications of
  the Astronomical Society of the Pacific, 131, 115003, aDS Bibcode:
  2019PASP..131k5003A

\bibitem[{Airapetian {et~al.}(2017)Airapetian, Glocer, Khazanov, Loyd, France,
  Sojka, Danchi, \& Liemohn}]{airapetian_how_2017}
Airapetian, V.~S., Glocer, A., Khazanov, G.~V., {et~al.} 2017, The
  Astrophysical Journal, 836, L3, aDS Bibcode: 2017ApJ...836L...3A

\bibitem[{Allart {et~al.}(2017)Allart, Lovis, Pino, Wyttenbach, Ehrenreich, \&
  Pepe}]{allart_search_2017}
Allart, R., Lovis, C., Pino, L., {et~al.} 2017, Astronomy and Astrophysics,
  606, A144, aDS Bibcode: 2017A\&A...606A.144A

\bibitem[{Apai {et~al.}(2018)Apai, Rackham, Giampapa, Angerhausen, Teske,
  Barstow, Carone, Cegla, Domagal-Goldman, Espinoza, Giles, Gully-Santiago,
  Haywood, Hu, Jordan, Kreidberg, Line, Llama, López-Morales, Marley, \&
  de~Wit}]{apai_understanding_2018}
Apai, D., Rackham, B.~V., Giampapa, M.~S., {et~al.} 2018, Understanding
  {Stellar} {Contamination} in {Exoplanet} {Transmission} {Spectra} as an
  {Essential} {Step} in {Small} {Planet} {Characterization}, arXiv:1803.08708
  [astro-ph]

\bibitem[{Aronson \& Waldén(2015)}]{aronson_using_2015}
Aronson, E. \& Waldén, P. 2015, Astronomy and Astrophysics, 578, A133, aDS
  Bibcode: 2015A\&A...578A.133A

\bibitem[{Artigau {et~al.}(2014)Artigau, Kouach, Donati, Doyon, Delfosse,
  Baratchart, Lacombe, Moutou, Rabou, Parès, Micheau, Thibault, Reshetov,
  Dubois, Hernandez, Vallée, Wang, Dolon, Pepe, Bouchy, Striebig, Hénault,
  Loop, Saddlemyer, Barrick, Vermeulen, Dupieux, Hébrard, Boisse, Martioli,
  Alencar, do~Nascimento, \& Figueira}]{artigau_spirou_2014}
Artigau, E., Kouach, D., Donati, J.-F., {et~al.} 2014, Conference Name:
  Ground-based and Airborne Instrumentation for Astronomy V, 9147, 914715,
  conference Name: Ground-based and Airborne Instrumentation for Astronomy V
  Place: eprint: arXiv:1406.6992 ADS Bibcode: 2014SPIE.9147E..15A

\bibitem[{{Astropy Collaboration} {et~al.}(2013){Astropy Collaboration},
  Robitaille, Tollerud, Greenfield, Droettboom, Bray, Aldcroft, Davis,
  Ginsburg, Price-Whelan, Kerzendorf, Conley, Crighton, Barbary, Muna,
  Ferguson, Grollier, Parikh, Nair, Unther, Deil, Woillez, Conseil, Kramer,
  Turner, Singer, Fox, Weaver, Zabalza, Edwards, Azalee~Bostroem, Burke, Casey,
  Crawford, Dencheva, Ely, Jenness, Labrie, Lim, Pierfederici, Pontzen, Ptak,
  Refsdal, Servillat, \& Streicher}]{astropy_collaboration_astropy_2013}
{Astropy Collaboration}, Robitaille, T.~P., Tollerud, E.~J., {et~al.} 2013,
  Astronomy and Astrophysics, 558, A33, aDS Bibcode: 2013A\&A...558A..33A

\bibitem[{Astudillo-Defru \& Rojo(2013)}]{astudillo-defru_ground-based_2013}
Astudillo-Defru, N. \& Rojo, P. 2013, Astronomy and Astrophysics, 557, A56, aDS
  Bibcode: 2013A\&A...557A..56A

\bibitem[{Bakos {et~al.}(2004)Bakos, Noyes, Kovács, Stanek, Sasselov, \&
  Domsa}]{bakos_wide-field_2004}
Bakos, G., Noyes, R.~W., Kovács, G., {et~al.} 2004, Publications of the
  Astronomical Society of the Pacific, 116, 266, aDS Bibcode:
  2004PASP..116..266B

\bibitem[{Bakos {et~al.}(2013)Bakos, Csubry, Penev, Bayliss, Jordán, Afonso,
  Hartman, Henning, Kovács, Noyes, Béky, Suc, Csák, Rabus, Lázár, Papp,
  Sári, Conroy, Zhou, Sackett, Schmidt, Mancini, Sasselov, \&
  Ueltzhoeffer}]{bakos_hatsouth_2013}
Bakos, G.~A., Csubry, Z., Penev, K., {et~al.} 2013, Publications of the
  Astronomical Society of the Pacific, 125, 154, aDS Bibcode:
  2013PASP..125..154B

\bibitem[{Barclay {et~al.}(2021)Barclay, Kostov, Colón, Quintana, Schlieder,
  Louie, Gilbert, \& Mullally}]{barclay_stellar_2021}
Barclay, T., Kostov, V.~B., Colón, K.~D., {et~al.} 2021, The Astronomical
  Journal, 162, 300, aDS Bibcode: 2021AJ....162..300B

\bibitem[{Bean {et~al.}(2010)Bean, Miller-Ricci~Kempton, \&
  Homeier}]{bean_ground-based_2010}
Bean, J.~L., Miller-Ricci~Kempton, E., \& Homeier, D. 2010, Nature, 468, 669,
  aDS Bibcode: 2010Natur.468..669B

\bibitem[{Benneke {et~al.}(2019)Benneke, Wong, Piaulet, Knutson, Lothringer,
  Morley, Crossfield, Gao, Greene, Dressing, Dragomir, Howard, McCullough,
  Kempton, Fortney, \& Fraine}]{benneke_water_2019}
Benneke, B., Wong, I., Piaulet, C., {et~al.} 2019, The Astrophysical Journal,
  887, L14, aDS Bibcode: 2019ApJ...887L..14B

\bibitem[{Birkby(2018)}]{birkby_exoplanet_2018}
Birkby, J.~L. 2018, Exoplanet {Atmospheres} at {High} {Spectral} {Resolution},
  arXiv:1806.04617 [astro-ph]

\bibitem[{Birkby {et~al.}(2013)Birkby, de~Kok, Brogi, de~Mooij, Schwarz,
  Albrecht, \& Snellen}]{birkby_detection_2013}
Birkby, J.~L., de~Kok, R.~J., Brogi, M., {et~al.} 2013, Monthly Notices of the
  Royal Astronomical Society, 436, L35, aDS Bibcode: 2013MNRAS.436L..35B

\bibitem[{Birkby {et~al.}(2017)Birkby, Kok, Brogi, Schwarz, \&
  Snellen}]{birkby_discovery_2017}
Birkby, J.~L., Kok, R. J.~d., Brogi, M., Schwarz, H., \& Snellen, I. A.~G.
  2017, The Astronomical Journal, 153, 138, publisher: American Astronomical
  Society

\bibitem[{Birkmann {et~al.}(2022)Birkmann, Ferruit, Giardino, Nielsen,
  García~Muñoz, Kendrew, Rauscher, Beck, Keyes, Valenti, Jakobsen, Dorner,
  Alves~de Oliveira, Arribas, Böker, Bunker, Charlot, de~Marchi, Kumari,
  López-Caniego, Lützgendorf, Maiolino, Manjavacas, Marston, Moseley,
  Prizkal, Proffitt, Rawle, Rix, te~Plate, Sabbi, Sirianni, Willott, \&
  Zeidler}]{birkmann_near-infrared_2022}
Birkmann, S.~M., Ferruit, P., Giardino, G., {et~al.} 2022, Astronomy and
  Astrophysics, 661, A83, aDS Bibcode: 2022A\&A...661A..83B

\bibitem[{Borucki {et~al.}(2010)Borucki, Koch, Basri, Batalha, Brown, Caldwell,
  Caldwell, Christensen-Dalsgaard, Cochran, DeVore, Dunham, Dupree, Gautier,
  Geary, Gilliland, Gould, Howell, Jenkins, Kondo, Latham, Marcy, Meibom,
  Kjeldsen, Lissauer, Monet, Morrison, Sasselov, Tarter, Boss, Brownlee, Owen,
  Buzasi, Charbonneau, Doyle, Fortney, Ford, Holman, Seager, Steffen, Welsh,
  Rowe, Anderson, Buchhave, Ciardi, Walkowicz, Sherry, Horch, Isaacson,
  Everett, Fischer, Torres, Johnson, Endl, MacQueen, Bryson, Dotson, Haas,
  Kolodziejczak, Van~Cleve, Chandrasekaran, Twicken, Quintana, Clarke, Allen,
  Li, Wu, Tenenbaum, Verner, Bruhweiler, Barnes, \& Prsa}]{borucki_kepler_2010}
Borucki, W.~J., Koch, D., Basri, G., {et~al.} 2010, Science, 327, 977, aDS
  Bibcode: 2010Sci...327..977B

\bibitem[{Boucher {et~al.}(2023)Boucher, Lafreniére, Pelletier,
  Darveau-Bernier, Radica, Allart, Artigau, Cook, Debras, Doyon, Gaidos,
  Benneke, Cadieux, Carmona, Cloutier, Cortés-Zuleta, Cowan, Delfosse, Donati,
  Fouqué, Forveille, Grankin, Hébrard, Martins, Martioli, Masson, \&
  Vinatier}]{boucher_co_2023}
Boucher, A., Lafreniére, D., Pelletier, S., {et~al.} 2023, Monthly Notices of
  the Royal Astronomical Society, 522, 5062, aDS Bibcode: 2023MNRAS.522.5062B

\bibitem[{Bouchy {et~al.}(2005)Bouchy, Udry, Mayor, Moutou, Pont, Iribarne,
  da~Silva, Ilovaisky, Queloz, Santos, Ségransan, \&
  Zucker}]{bouchy_elodie_2005}
Bouchy, F., Udry, S., Mayor, M., {et~al.} 2005, Astronomy and Astrophysics,
  444, L15, aDS Bibcode: 2005A\&A...444L..15B

\bibitem[{Brogi {et~al.}(2016)Brogi, de~Kok, Albrecht, Snellen, Birkby, \&
  Schwarz}]{brogi_rotation_2016}
Brogi, M., de~Kok, R.~J., Albrecht, S., {et~al.} 2016, The Astrophysical
  Journal, 817, 106, aDS Bibcode: 2016ApJ...817..106B

\bibitem[{Brogi {et~al.}(2018)Brogi, Giacobbe, Guilluy, Kok, Sozzetti, Mancini,
  \& Bonomo}]{brogi_exoplanet_2018}
Brogi, M., Giacobbe, P., Guilluy, G., {et~al.} 2018, Astronomy \& Astrophysics,
  615, A16, publisher: EDP Sciences

\bibitem[{Brogi {et~al.}(2017)Brogi, Line, Bean, Désert, \&
  Schwarz}]{brogi_framework_2017}
Brogi, M., Line, M., Bean, J., Désert, J.-M., \& Schwarz, H. 2017, The
  Astrophysical Journal, 839, L2, publisher: American Astronomical Society

\bibitem[{Brogi \& Line(2019)}]{brogi_retrieving_2019}
Brogi, M. \& Line, M.~R. 2019, The Astronomical Journal, 157, 114, aDS Bibcode:
  2019AJ....157..114B

\bibitem[{Cabot {et~al.}(2019)Cabot, Madhusudhan, Hawker, \&
  Gandhi}]{cabot_robustness_2019}
Cabot, S. H.~C., Madhusudhan, N., Hawker, G.~A., \& Gandhi, S. 2019, Monthly
  Notices of the Royal Astronomical Society, 482, 4422, aDS Bibcode:
  2019MNRAS.482.4422C

\bibitem[{Charbonneau {et~al.}(2002)Charbonneau, Brown, Noyes, \&
  Gilliland}]{charbonneau_detection_2002}
Charbonneau, D., Brown, T.~M., Noyes, R.~W., \& Gilliland, R.~L. 2002, The
  Astrophysical Journal, 568, 377, arXiv:astro-ph/0111544

\bibitem[{Cheverall {et~al.}(2023)Cheverall, Madhusudhan, \&
  Holmberg}]{cheverall_robustness_2023}
Cheverall, C.~J., Madhusudhan, N., \& Holmberg, M. 2023, Monthly Notices of the
  Royal Astronomical Society, 522, 661

\bibitem[{Cutri {et~al.}(2003)Cutri, Skrutskie, van Dyk, Beichman, Carpenter,
  Chester, Cambresy, Evans, Fowler, Gizis, Howard, Huchra, Jarrett, Kopan,
  Kirkpatrick, Light, Marsh, McCallon, Schneider, Stiening, Sykes, Weinberg,
  Wheaton, Wheelock, \& Zacarias}]{cutri_vizier_2003}
Cutri, R.~M., Skrutskie, M.~F., van Dyk, S., {et~al.} 2003, VizieR Online Data
  Catalog, II/246, aDS Bibcode: 2003yCat.2246....0C

\bibitem[{Czesla {et~al.}(2019)Czesla, Schröter, Schneider, Huber, Pfeifer,
  Andreasen, \& Zechmeister}]{czesla_pya_2019}
Czesla, S., Schröter, S., Schneider, C.~P., {et~al.} 2019, {PyA}: {Python}
  astronomy-related packages, pages: ascl:1906.010 \_eprint: 1906.010

\bibitem[{Di~Gloria {et~al.}(2015)Di~Gloria, Snellen, \&
  Albrecht}]{di_gloria_using_2015}
Di~Gloria, E., Snellen, I. A.~G., \& Albrecht, S. 2015, Astronomy and
  Astrophysics, 580, A84, aDS Bibcode: 2015A\&A...580A..84D

\bibitem[{Diamond-Lowe {et~al.}(2018)Diamond-Lowe, Berta-Thompson, Charbonneau,
  \& Kempton}]{diamond-lowe_ground-based_2018}
Diamond-Lowe, H., Berta-Thompson, Z., Charbonneau, D., \& Kempton, E. M.~R.
  2018, The Astronomical Journal, 156, 42, aDS Bibcode: 2018AJ....156...42D

\bibitem[{Diamond-Lowe {et~al.}(2023)Diamond-Lowe, Mendonça, Charbonneau, \&
  Buchhave}]{diamond-lowe_ground-based_2023}
Diamond-Lowe, H., Mendonça, J.~M., Charbonneau, D., \& Buchhave, L.~A. 2023,
  The Astronomical Journal, 165, 169, aDS Bibcode: 2023AJ....165..169D

\bibitem[{Ding \& Wordsworth(2019)}]{ding_new_2019}
Ding, F. \& Wordsworth, R.~D. 2019, The Astrophysical Journal, 878, 117,
  publisher: The American Astronomical Society

\bibitem[{Dorn {et~al.}(2023)Dorn, Bristow, Smoker, Rodler, Lavail, Accardo,
  Ancker, Baade, Baruffolo, Courtney-Barrer, Blanco, Brucalassi, Cumani,
  Follert, Haimerl, Hatzes, Haug, Heiter, Hinterschuster, Hubin, Ives, Jung,
  Jones, Kaeufl, Kirchbauer, Klein, Kochukhov, Korhonen, Köhler, Lizon, Moins,
  Molina-Conde, Marquart, Neeser, Oliva, Pallanca, Pasquini, Paufique,
  Piskunov, Reiners, Schneller, Schmutzer, Seemann, Slumstrup, Smette,
  Stegmeier, Stempels, Tordo, Valenti, Valenzuela, Vernet, Vinther, \&
  Wehrhahn}]{dorn_crires_2023}
Dorn, R.~J., Bristow, P., Smoker, J.~V., {et~al.} 2023, Astronomy \&
  Astrophysics, 671, A24, publisher: EDP Sciences

\bibitem[{Ehrenreich {et~al.}(2020)Ehrenreich, Lovis, Allart, Zapatero~Osorio,
  Pepe, Cristiani, Rebolo, Santos, Borsa, Demangeon, Dumusque,
  González~Hernández, Casasayas-Barris, Ségransan, Sousa, Abreu, Adibekyan,
  Affolter, Allende~Prieto, Alibert, Aliverti, Alves, Amate, Avila, Baldini,
  Bandy, Benz, Bianco, Bolmont, Bouchy, Bourrier, Broeg, Cabral, Calderone,
  Pallé, Cegla, Cirami, Coelho, Conconi, Coretti, Cumani, Cupani, Dekker,
  Delabre, Deiries, D'Odorico, Di~Marcantonio, Figueira, Fragoso, Genolet,
  Genoni, Génova~Santos, Hara, Hughes, Iwert, Kerber, Knudstrup, Landoni,
  Lavie, Lizon, Lendl, Lo~Curto, Maire, Manescau, Martins, Mégevand, Mehner,
  Micela, Modigliani, Molaro, Monteiro, Monteiro, Moschetti, Müller, Nunes,
  Oggioni, Oliveira, Pariani, Pasquini, Poretti, Rasilla, Redaelli, Riva,
  Santana~Tschudi, Santin, Santos, Segovia~Milla, Seidel, Sosnowska, Sozzetti,
  Spanò, Suárez~Mascareño, Tabernero, Tenegi, Udry, Zanutta, \&
  Zerbi}]{ehrenreich_nightside_2020}
Ehrenreich, D., Lovis, C., Allart, R., {et~al.} 2020, Nature, 580, 597

\bibitem[{Fauchez {et~al.}(2022)Fauchez, Villanueva, Sergeev, Turbet, Boutle,
  Tsigaridis, Way, Wolf, Domagal-Goldman, Forget, Haqq-Misra, Kopparapu,
  Manners, \& Mayne}]{fauchez_trappist-1_2022}
Fauchez, T.~J., Villanueva, G.~L., Sergeev, D.~E., {et~al.} 2022, The Planetary
  Science Journal, 3, 213, arXiv:2109.11460 [astro-ph, physics:physics]

\bibitem[{Fortney {et~al.}(2021)Fortney, Dawson, \& Komacek}]{fortney_hot_2021}
Fortney, J.~J., Dawson, R.~I., \& Komacek, T.~D. 2021, Journal of Geophysical
  Research: Planets, 126, e2020JE006629, \_eprint:
  https://onlinelibrary.wiley.com/doi/pdf/10.1029/2020JE006629

\bibitem[{{Gaia Collaboration} {et~al.}(2018){Gaia Collaboration}, Brown,
  Vallenari, Prusti, de~Bruijne, Babusiaux, Bailer-Jones, Biermann, Evans,
  Eyer, Jansen, Jordi, Klioner, Lammers, Lindegren, Luri, Mignard, Panem,
  Pourbaix, Randich, Sartoretti, Siddiqui, Soubiran, van Leeuwen, Walton,
  Arenou, Bastian, Cropper, Drimmel, Katz, Lattanzi, Bakker, Cacciari,
  Castañeda, Chaoul, Cheek, De~Angeli, Fabricius, Guerra, Holl, Masana,
  Messineo, Mowlavi, Nienartowicz, Panuzzo, Portell, Riello, Seabroke, Tanga,
  Thévenin, Gracia-Abril, Comoretto, Garcia-Reinaldos, Teyssier, Altmann,
  Andrae, Audard, Bellas-Velidis, Benson, Berthier, Blomme, Burgess, Busso,
  Carry, Cellino, Clementini, Clotet, Creevey, Davidson, De~Ridder, Delchambre,
  Dell'Oro, Ducourant, Fernández-Hernández, Fouesneau, Frémat, Galluccio,
  García-Torres, González-Núñez, González-Vidal, Gosset, Guy, Halbwachs,
  Hambly, Harrison, Hernández, Hestroffer, Hodgkin, Hutton, Jasniewicz,
  Jean-Antoine-Piccolo, Jordan, Korn, Krone-Martins, Lanzafame, Lebzelter,
  Löffler, Manteiga, Marrese, Martín-Fleitas, Moitinho, Mora, Muinonen,
  Osinde, Pancino, Pauwels, Petit, Recio-Blanco, Richards, Rimoldini, Robin,
  Sarro, Siopis, Smith, Sozzetti, Süveges, Torra, van Reeven, Abbas,
  Abreu~Aramburu, Accart, Aerts, Altavilla, Álvarez, Alvarez, Alves, Anderson,
  Andrei, Anglada~Varela, Antiche, Antoja, Arcay, Astraatmadja, Bach, Baker,
  Balaguer-Núñez, Balm, Barache, Barata, Barbato, Barblan, Barklem, Barrado,
  Barros, Barstow, Bartholomé~Muñoz, Bassilana, Becciani, Bellazzini,
  Berihuete, Bertone, Bianchi, Bienaymé, Blanco-Cuaresma, Boch, Boeche,
  Bombrun, Borrachero, Bossini, Bouquillon, Bourda, Bragaglia, Bramante,
  Breddels, Bressan, Brouillet, Brüsemeister, Brugaletta, Bucciarelli,
  Burlacu, Busonero, Butkevich, Buzzi, Caffau, Cancelliere, Cannizzaro,
  Cantat-Gaudin, Carballo, Carlucci, Carrasco, Casamiquela, Castellani,
  Castro-Ginard, Charlot, Chemin, Chiavassa, Cocozza, Costigan, Cowell, Crifo,
  Crosta, Crowley, Cuypers, Dafonte, Damerdji, Dapergolas, David, David,
  de~Laverny, De~Luise, De~March, de~Martino, de~Souza, de~Torres, Debosscher,
  del Pozo, Delbo, Delgado, Delgado, Di~Matteo, Diakite, Diener, Distefano,
  Dolding, Drazinos, Durán, Edvardsson, Enke, Eriksson, Esquej,
  Eynard~Bontemps, Fabre, Fabrizio, Faigler, Falcão, Farràs~Casas, Federici,
  Fedorets, Fernique, Figueras, Filippi, Findeisen, Fonti, Fraile, Fraser,
  Frézouls, Gai, Galleti, Garabato, García-Sedano, Garofalo, Garralda, Gavel,
  Gavras, Gerssen, Geyer, Giacobbe, Gilmore, Girona, Giuffrida, Glass, Gomes,
  Granvik, Gueguen, Guerrier, Guiraud, Gutiérrez-Sánchez, Haigron,
  Hatzidimitriou, Hauser, Haywood, Heiter, Helmi, Heu, Hilger, Hobbs, Hofmann,
  Holland, Huckle, Hypki, Icardi, Janßen, Jevardat~de Fombelle, Jonker,
  Juhász, Julbe, Karampelas, Kewley, Klar, Kochoska, Kohley, Kolenberg,
  Kontizas, Kontizas, Koposov, Kordopatis, Kostrzewa-Rutkowska, Koubsky,
  Lambert, Lanza, Lasne, Lavigne, Le~Fustec, Le~Poncin-Lafitte, Lebreton,
  Leccia, Leclerc, Lecoeur-Taibi, Lenhardt, Leroux, Liao, Licata, Lindstrøm,
  Lister, Livanou, Lobel, López, Managau, Mann, Mantelet, Marchal, Marchant,
  Marconi, Marinoni, Marschalkó, Marshall, Martino, Marton, Mary, Massari,
  Matijevič, Mazeh, McMillan, Messina, Michalik, Millar, Molina, Molinaro,
  Molnár, Montegriffo, Mor, Morbidelli, Morel, Morris, Mulone, Muraveva,
  Musella, Nelemans, Nicastro, Noval, O'Mullane, Ordénovic, Ordóñez-Blanco,
  Osborne, Pagani, Pagano, Pailler, Palacin, Palaversa, Panahi, Pawlak,
  Piersimoni, Pineau, Plachy, Plum, Poggio, Poujoulet, Prša, Pulone, Racero,
  Ragaini, Rambaux, Ramos-Lerate, Regibo, Reylé, Riclet, Ripepi, Riva, Rivard,
  Rixon, Roegiers, Roelens, Romero-Gómez, Rowell, Royer, Ruiz-Dern, Sadowski,
  Sagristà~Sellés, Sahlmann, Salgado, Salguero, Sanna, Santana-Ros, Sarasso,
  Savietto, Schultheis, Sciacca, Segol, Segovia, Ségransan, Shih, Siltala,
  Silva, Smart, Smith, Solano, Solitro, Sordo, Soria~Nieto, Souchay, Spagna,
  Spoto, Stampa, Steele, Steidelmüller, Stephenson, Stoev, Suess, Surdej,
  Szabados, Szegedi-Elek, Tapiador, Taris, Tauran, Taylor, Teixeira, Terrett,
  Teyssandier, Thuillot, Titarenko, Torra~Clotet, Turon, Ulla, Utrilla, Uzzi,
  Vaillant, Valentini, Valette, van Elteren, Van~Hemelryck, van Leeuwen,
  Vaschetto, Vecchiato, Veljanoski, Viala, Vicente, Vogt, von Essen, Voss,
  Votruba, Voutsinas, Walmsley, Weiler, Wertz, Wevers, Wyrzykowski, Yoldas,
  Žerjal, Ziaeepour, Zorec, Zschocke, Zucker, Zurbach, \&
  Zwitter}]{gaia_collaboration_gaia_2018}
{Gaia Collaboration}, Brown, A. G.~A., Vallenari, A., {et~al.} 2018, Astronomy
  and Astrophysics, 616, A1, aDS Bibcode: 2018A\&A...616A...1G

\bibitem[{Gandhi {et~al.}(2023)Gandhi, Kesseli, Zhang, Louca, Snellen, Brogi,
  Miguel, Casasayas-Barris, Pelletier, Landman, Maguire, \&
  Gibson}]{gandhi_retrieval_2023}
Gandhi, S., Kesseli, A., Zhang, Y., {et~al.} 2023, Retrieval survey of metals
  in six ultra-hot {Jupiters}: {Trends} in chemistry, rain-out, ionisation and
  atmospheric dynamics, Tech. rep., publication Title: arXiv e-prints ADS
  Bibcode: 2023arXiv230517228G Type: article

\bibitem[{Gao {et~al.}(2021)Gao, Wakeford, Moran, \&
  Parmentier}]{gao_aerosols_2021}
Gao, P., Wakeford, H.~R., Moran, S.~E., \& Parmentier, V. 2021, Journal of
  Geophysical Research (Planets), 126, e06655, aDS Bibcode: 2021JGRE..12606655G

\bibitem[{Giacobbe {et~al.}(2021)Giacobbe, Brogi, Gandhi, Cubillos, Bonomo,
  Sozzetti, Fossati, Guilluy, Carleo, Rainer, Harutyunyan, Borsa, Pino,
  Nascimbeni, Benatti, Biazzo, Bignamini, Chubb, Claudi, Cosentino, Covino,
  Damasso, Desidera, Fiorenzano, Ghedina, Lanza, Leto, Maggio, Malavolta,
  Maldonado, Micela, Molinari, Pagano, Pedani, Piotto, Poretti, Scandariato,
  Yurchenko, Fantinel, Galli, Lodi, Sanna, \& Tozzi}]{giacobbe_five_2021}
Giacobbe, P., Brogi, M., Gandhi, S., {et~al.} 2021, Nature, 592, 205, number:
  7853 Publisher: Nature Publishing Group

\bibitem[{Gibson {et~al.}(2020)Gibson, Merritt, Nugroho, Cubillos, de~Mooij,
  Mikal-Evans, Fossati, Lothringer, Nikolov, Sing, Spake, Watson, \&
  Wilson}]{gibson_detection_2020}
Gibson, N.~P., Merritt, S., Nugroho, S.~K., {et~al.} 2020, Monthly Notices of
  the Royal Astronomical Society, 493, 2215, aDS Bibcode: 2020MNRAS.493.2215G

\bibitem[{Hardegree-Ullman {et~al.}(2019)Hardegree-Ullman, Cushing, Muirhead,
  \& Christiansen}]{hardegree-ullman_kepler_2019}
Hardegree-Ullman, K.~K., Cushing, M.~C., Muirhead, P.~S., \& Christiansen,
  J.~L. 2019, The Astronomical Journal, 158, 75, aDS Bibcode:
  2019AJ....158...75H

\bibitem[{Harris {et~al.}(2020)Harris, Millman, van~der Walt, Gommers,
  Virtanen, Cournapeau, Wieser, Taylor, Berg, Smith, Kern, Picus, Hoyer, van
  Kerkwijk, Brett, Haldane, del Río, Wiebe, Peterson, Gérard-Marchant,
  Sheppard, Reddy, Weckesser, Abbasi, Gohlke, \& Oliphant}]{harris_array_2020}
Harris, C.~R., Millman, K.~J., van~der Walt, S.~J., {et~al.} 2020, Nature, 585,
  357, number: 7825 Publisher: Nature Publishing Group

\bibitem[{Hawker {et~al.}(2018)Hawker, Madhusudhan, Cabot, \&
  Gandhi}]{hawker_evidence_2018}
Hawker, G.~A., Madhusudhan, N., Cabot, S. H.~C., \& Gandhi, S. 2018, The
  Astrophysical Journal Letters, 863, L11, publisher: The American Astronomical
  Society

\bibitem[{Helling(2019)}]{helling_exoplanet_2019}
Helling, C. 2019, Annual Review of Earth and Planetary Sciences, 47, 583,
  \_eprint: https://doi.org/10.1146/annurev-earth-053018-060401

\bibitem[{Heng \& Showman(2015)}]{heng_atmospheric_2015}
Heng, K. \& Showman, A.~P. 2015, Annual Review of Earth and Planetary Sciences,
  43, 509, \_eprint: https://doi.org/10.1146/annurev-earth-060614-105146

\bibitem[{Hoeijmakers {et~al.}(2018)Hoeijmakers, Ehrenreich, Heng, Kitzmann,
  Grimm, Allart, Deitrick, Wyttenbach, Oreshenko, Pino, Rimmer, Molinari, \&
  Di~Fabrizio}]{hoeijmakers_atomic_2018}
Hoeijmakers, H.~J., Ehrenreich, D., Heng, K., {et~al.} 2018, Nature, 560, 453,
  aDS Bibcode: 2018Natur.560..453H

\bibitem[{Howell {et~al.}(2014)Howell, Sobeck, Haas, Still, Barclay, Mullally,
  Troeltzsch, Aigrain, Bryson, Caldwell, Chaplin, Cochran, Huber, Marcy,
  Miglio, Najita, Smith, Twicken, \& Fortney}]{howell_k2_2014}
Howell, S.~B., Sobeck, C., Haas, M., {et~al.} 2014, Publications of the
  Astronomical Society of the Pacific, 126, 398, aDS Bibcode:
  2014PASP..126..398H

\bibitem[{Hunter(2007)}]{hunter_matplotlib_2007}
Hunter, J.~D. 2007, Computing in Science \& Engineering, 9, 90, conference
  Name: Computing in Science \& Engineering

\bibitem[{Husser {et~al.}(2013)Husser, Berg, Dreizler, Homeier, Reiners,
  Barman, \& Hauschildt}]{husser_new_2013}
Husser, T.-O., Berg, S. W.-v., Dreizler, S., {et~al.} 2013, Astronomy \&
  Astrophysics, 553, A6, publisher: EDP Sciences

\bibitem[{Jakobsen {et~al.}(2022)Jakobsen, Ferruit, Oliveira, Arribas,
  Bagnasco, Barho, Beck, Birkmann, Böker, Bunker, Charlot, Jong, Marchi,
  Ehrenwinkler, Falcolini, Fels, Franx, Franz, Funke, Giardino, Gnata, Holota,
  Honnen, Jensen, Jentsch, Johnson, Jollet, Karl, Kling, Köhler, Kolm, Kumari,
  Lander, Lemke, López-Caniego, Lützgendorf, Maiolino, Manjavacas, Marston,
  Maschmann, Maurer, Messerschmidt, Moseley, Mosner, Mott, Muzerolle, Pirzkal,
  Pittet, Plitzke, Posselt, Rapp, Rauscher, Rawle, Rix, Rödel, Rumler, Sabbi,
  Salvignol, Schmid, Sirianni, Smith, Strada, Plate, Valenti, Wettemann, Wiehe,
  Wiesmayer, Willott, Wright, Zeidler, \& Zincke}]{jakobsen_near-infrared_2022}
Jakobsen, P., Ferruit, P., Oliveira, C. A.~d., {et~al.} 2022, Astronomy \&
  Astrophysics, 661, A80, publisher: EDP Sciences

\bibitem[{Jehin {et~al.}(2011)Jehin, Gillon, Queloz, Magain, Manfroid, Chantry,
  Lendl, Hutsemékers, \& Udry}]{jehin_trappist_2011}
Jehin, E., Gillon, M., Queloz, D., {et~al.} 2011, The Messenger, 145, 2, aDS
  Bibcode: 2011Msngr.145....2J

\bibitem[{{JWST Transiting Exoplanet Community Early Release Science Team}
  {et~al.}(2023){JWST Transiting Exoplanet Community Early Release Science
  Team}, Ahrer, Alderson, Batalha, Batalha, Bean, Beatty, Bell, Benneke,
  Berta-Thompson, Carter, Crossfield, Espinoza, Feinstein, Fortney, Gibson,
  Goyal, Kempton, Kirk, Kreidberg, López-Morales, Line, Lothringer, Moran,
  Mukherjee, Ohno, Parmentier, Piaulet, Rustamkulov, Schlawin, Sing, Stevenson,
  Wakeford, Allen, Birkmann, Brande, Crouzet, Cubillos, Damiano, Désert, Gao,
  Harrington, Hu, Kendrew, Knutson, Lagage, Leconte, Lendl, MacDonald, May,
  Miguel, Molaverdikhani, Moses, Murray, Nehring, Nikolov, Petit dit de~la
  Roche, Radica, Roy, Stassun, Taylor, Waalkes, Wachiraphan, Welbanks,
  Wheatley, Aggarwal, Alam, Banerjee, Barstow, Blecic, Casewell, Changeat,
  Chubb, Colón, Coulombe, Daylan, de~Val-Borro, Decin, Dos~Santos, Flagg,
  France, Fu, García~Muñoz, Gizis, Glidden, Grant, Heng, Henning, Hong,
  Inglis, Iro, Kataria, Komacek, Krick, Lee, Lewis, Lillo-Box, Lustig-Yaeger,
  Mancini, Mandell, Mansfield, Marley, Mikal-Evans, Morello, Nixon,
  Ortiz~Ceballos, Piette, Powell, Rackham, Ramos-Rosado, Rauscher, Redfield,
  Rogers, Roman, Roudier, Scarsdale, Shkolnik, Southworth, Spake, Steinrueck,
  Tan, Teske, Tremblin, Tsai, Tucker, Turner, Valenti, Venot, Waldmann,
  Wallack, Zhang, \&
  Zieba}]{jwst_transiting_exoplanet_community_early_release_science_team_identification_2023}
{JWST Transiting Exoplanet Community Early Release Science Team}, Ahrer, E.-M.,
  Alderson, L., {et~al.} 2023, Nature, 614, 649, aDS Bibcode:
  2023Natur.614..649J

\bibitem[{Kaltenegger \& Traub(2009)}]{kaltenegger_transits_2009}
Kaltenegger, L. \& Traub, W.~A. 2009, The Astrophysical Journal, 698, 519,
  publisher: The American Astronomical Society

\bibitem[{Kanodia {et~al.}(2019)Kanodia, Wolfgang, Stefansson, Ning, \&
  Mahadevan}]{kanodia_mass-radius_2019}
Kanodia, S., Wolfgang, A., Stefansson, G.~K., Ning, B., \& Mahadevan, S. 2019,
  The Astrophysical Journal, 882, 38, aDS Bibcode: 2019ApJ...882...38K

\bibitem[{Kausch {et~al.}(2015)Kausch, Noll, Smette, Kimeswenger, Barden,
  Szyszka, Jones, Sana, Horst, \& Kerber}]{kausch_molecfit_2015}
Kausch, W., Noll, S., Smette, A., {et~al.} 2015, Astronomy \& Astrophysics,
  576, A78, publisher: EDP Sciences

\bibitem[{Kesseli {et~al.}(2022)Kesseli, Snellen, Casasayas-Barris, Mollière,
  \& Sánchez-López}]{kesseli_atomic_2022}
Kesseli, A.~Y., Snellen, I. A.~G., Casasayas-Barris, N., Mollière, P., \&
  Sánchez-López, A. 2022, The Astronomical Journal, 163, 107, publisher: The
  American Astronomical Society

\bibitem[{Kochukhov(2021)}]{kochukhov_magnetic_2021}
Kochukhov, O. 2021, Astronomy and Astrophysics Review, 29, 1, aDS Bibcode:
  2021A\&ARv..29....1K

\bibitem[{Kok {et~al.}(2014)Kok, Birkby, Brogi, Schwarz, Albrecht, de~Mooij, \&
  Snellen}]{kok_identifying_2014}
Kok, R. J.~d., Birkby, J., Brogi, M., {et~al.} 2014, Astronomy and
  Astrophysics, 561, A150

\bibitem[{Kornilov {et~al.}(2007)Kornilov, Tokovinin, Shatsky, Voziakova,
  Potanin, \& Safonov}]{kornilov_combined_2007}
Kornilov, V., Tokovinin, A., Shatsky, N., {et~al.} 2007, Monthly Notices of the
  Royal Astronomical Society, 382, 1268

\bibitem[{Kreidberg {et~al.}(2014)Kreidberg, Bean, Désert, Benneke, Deming,
  Stevenson, Seager, Berta-Thompson, Seifahrt, \&
  Homeier}]{kreidberg_clouds_2014}
Kreidberg, L., Bean, J.~L., Désert, J.-M., {et~al.} 2014, Nature, 505, 69, aDS
  Bibcode: 2014Natur.505...69K

\bibitem[{Kreidberg {et~al.}(2018)Kreidberg, Line, Thorngren, Morley, \&
  Stevenson}]{kreidberg_water_2018}
Kreidberg, L., Line, M.~R., Thorngren, D., Morley, C.~V., \& Stevenson, K.~B.
  2018, The Astrophysical Journal Letters, 858, L6, publisher: The American
  Astronomical Society

\bibitem[{Kunimoto \& Matthews(2020)}]{kunimoto_searching_2020}
Kunimoto, M. \& Matthews, J.~M. 2020, The Astronomical Journal, 159, 248,
  publisher: American Astronomical Society

\bibitem[{Lam {et~al.}(2017)Lam, Faedi, Brown, Anderson, Delrez, Gillon,
  Hébrard, Lendl, Mancini, Southworth, Smalley, Triaud, Turner, Hay,
  Armstrong, Barros, Bonomo, Bouchy, Boumis, Collier~Cameron, Doyle, Hellier,
  Henning, Jehin, King, Kirk, Louden, Maxted, McCormac, Osborn, Palle, Pepe,
  Pollacco, Prieto-Arranz, Queloz, Rey, Ségransan, Udry, Walker, West, \&
  Wheatley}]{lam_dense_2017}
Lam, K. W.~F., Faedi, F., Brown, D. J.~A., {et~al.} 2017, Astronomy and
  Astrophysics, 599, A3, aDS Bibcode: 2017A\&A...599A...3L

\bibitem[{Lesjak {et~al.}(2023)Lesjak, Nortmann, Yan, Cont, Reiners, Piskunov,
  Hatzes, Boldt-Christmas, Czesla, Heiter, Kochukhov, Lavail, Nagel, Rains,
  Rengel, Rodler, Seemann, \& Shulyak}]{lesjak_retrieval_2023}
Lesjak, F., Nortmann, L., Yan, F., {et~al.} 2023, Astronomy and Astrophysics,
  678, A23, aDS Bibcode: 2023A\&A...678A..23L

\bibitem[{Libby-Roberts {et~al.}(2023)Libby-Roberts, Schutte, Hebb, Kanodia,
  Cañas, Stefánsson, Lin, Mahadevan, Parts, Powers, Wisniewski, Bender,
  Cochran, Diddams, Everett, Gupta, Halverson, Kobulnicky, Kowalski, Larsen,
  Monson, Ninan, Parker, Ramsey, Robertson, Schwab, Swaby, \&
  Terrien}]{libby-roberts_-depth_2023}
Libby-Roberts, J.~E., Schutte, M., Hebb, L., {et~al.} 2023, The Astronomical
  Journal, 165, 249, aDS Bibcode: 2023AJ....165..249L

\bibitem[{Loftus \& Wordsworth(2021)}]{loftus_physics_2021}
Loftus, K. \& Wordsworth, R. 2021, Journal of Geophysical Research: Planets,
  126, arXiv:2102.09570 [astro-ph, physics:physics]

\bibitem[{Lustig-Yaeger {et~al.}(2023)Lustig-Yaeger, Fu, May, Ceballos, Moran,
  Peacock, Stevenson, Kirk, López-Morales, MacDonald, Mayorga, Sing, Sotzen,
  Valenti, Redai, Alam, Batalha, Bennett, Gonzalez-Quiles, Kruse, Lothringer,
  Rustamkulov, \& Wakeford}]{lustig-yaeger_jwst_2023}
Lustig-Yaeger, J., Fu, G., May, E.~M., {et~al.} 2023, Nature Astronomy, aDS
  Bibcode: 2023NatAs.tmp..179L

\bibitem[{MacDonald \& Madhusudhan(2017)}]{macdonald_hd_2017}
MacDonald, R.~J. \& Madhusudhan, N. 2017, Monthly Notices of the Royal
  Astronomical Society, 469, 1979, aDS Bibcode: 2017MNRAS.469.1979M

\bibitem[{Maguire {et~al.}(2023)Maguire, Gibson, Nugroho, Ramkumar, Fortune,
  Merritt, \& de~Mooij}]{maguire_high-resolution_2023}
Maguire, C., Gibson, N.~P., Nugroho, S.~K., {et~al.} 2023, Monthly Notices of
  the Royal Astronomical Society, 519, 1030, aDS Bibcode: 2023MNRAS.519.1030M

\bibitem[{Maimone {et~al.}(2022)Maimone, Brogi, Chiavassa, van~den Ancker,
  Manara, Leconte, Gandhi, \& Pluriel}]{maimone_detecting_2022}
Maimone, M.~C., Brogi, M., Chiavassa, A., {et~al.} 2022, Detecting
  {H}\$\_2\${O} with {CRIRES}+: the case of {WASP}-20b, Tech. rep., publication
  Title: arXiv e-prints ADS Bibcode: 2022arXiv220911506M Type: article

\bibitem[{Mayor {et~al.}(2003)Mayor, Pepe, Queloz, Bouchy, Rupprecht, Lo~Curto,
  Avila, Benz, Bertaux, Bonfils, Dall, Dekker, Delabre, Eckert, Fleury,
  Gilliotte, Gojak, Guzman, Kohler, Lizon, Longinotti, Lovis, Megevand,
  Pasquini, Reyes, Sivan, Sosnowska, Soto, Udry, van Kesteren, Weber, \&
  Weilenmann}]{mayor_setting_2003}
Mayor, M., Pepe, F., Queloz, D., {et~al.} 2003, The Messenger, 114, 20, aDS
  Bibcode: 2003Msngr.114...20M

\bibitem[{McCloat {et~al.}(2021)McCloat, von Essen, \&
  Fieber-Beyer}]{mccloat_atmospheric_2021}
McCloat, S., von Essen, C., \& Fieber-Beyer, S. 2021, The Astronomical Journal,
  162, 132, aDS Bibcode: 2021AJ....162..132M

\bibitem[{McKinney(2010)}]{mckinney_data_2010}
McKinney, W. 2010, in Conference Proceedings: Python in Science Conference,
  Austin, Texas, 56--61

\bibitem[{Merritt {et~al.}(2021)Merritt, Gibson, Nugroho, de Mooij, Hooton,
  Lothringer, Matthews, Mikal-Evans, Nikolov, Sing, \&
  Watson}]{merritt_inventory_2021}
Merritt, S.~R., Gibson, N.~P., Nugroho, S.~K., {et~al.} 2021, Monthly Notices
  of the Royal Astronomical Society, 506, 3853

\bibitem[{Mollière {et~al.}(2019)Mollière, Wardenier, Boekel, Henning,
  Molaverdikhani, \& Snellen}]{molliere_petitradtrans_2019}
Mollière, P., Wardenier, J.~P., Boekel, R.~v., {et~al.} 2019, Astronomy \&
  Astrophysics, 627, A67, publisher: EDP Sciences

\bibitem[{Moran {et~al.}(2023)Moran, Stevenson, Sing, MacDonald, Kirk,
  Lustig-Yaeger, Peacock, Mayorga, Bennett, López-Morales, May, Rustamkulov,
  Valenti, Adams~Redai, Alam, Batalha, Fu, Gonzalez-Quiles, Highland, Kruse,
  Lothringer, Ortiz~Ceballos, Sotzen, \& Wakeford}]{moran_high_2023}
Moran, S.~E., Stevenson, K.~B., Sing, D.~K., {et~al.} 2023, The Astrophysical
  Journal, 948, L11, aDS Bibcode: 2023ApJ...948L..11M

\bibitem[{Morin {et~al.}(2010)Morin, Donati, Petit, Delfosse, Forveille, \&
  Jardine}]{morin_large-scale_2010}
Morin, J., Donati, J.~F., Petit, P., {et~al.} 2010, Monthly Notices of the
  Royal Astronomical Society, 407, 2269, aDS Bibcode: 2010MNRAS.407.2269M

\bibitem[{Mounzer {et~al.}(2022)Mounzer, Lovis, Seidel, Attia, Allart,
  Bourrier, Ehrenreich, Wyttenbach, Astudillo-Defru, Beatty, Cegla, Heng,
  Lavie, Lendl, Melo, Pepe, Pepper, Rodriguez, Ségransan, Udry, Linder, \&
  Sousa}]{mounzer_hot_2022}
Mounzer, D., Lovis, C., Seidel, J.~V., {et~al.} 2022, Astronomy and
  Astrophysics, 668, A1, aDS Bibcode: 2022A\&A...668A...1M

\bibitem[{Nikolov {et~al.}(2021)Nikolov, Maciejewski, Constantinou,
  Madhusudhan, Fortney, Smalley, Carter, Mooij, Drummond, Gibson, Helling,
  Mayne, Mikal-Evans, Sing, \& Wilson}]{nikolov_ground-based_2021}
Nikolov, N., Maciejewski, G., Constantinou, S., {et~al.} 2021, The Astronomical
  Journal, 162, 88, publisher: The American Astronomical Society

\bibitem[{Park {et~al.}(2014)Park, Jaffe, Yuk, Chun, Pak, Kim, Pavel, Lee, Oh,
  Jeong, Sim, Lee, Nguyen~Le, Strubhar, Gully-Santiago, Oh, Cha, Moon, Park,
  Brooks, Ko, Han, Nah, Hill, Lee, Barnes, Yu, Kaplan, Mace, Kim, Lee, Hwang,
  \& Park}]{park_design_2014}
Park, C., Jaffe, D.~T., Yuk, I.-S., {et~al.} 2014, Conference Name:
  Ground-based and Airborne Instrumentation for Astronomy V ADS Bibcode:
  2014SPIE.9147E..1DP, 9147, 91471D

\bibitem[{Pepe {et~al.}(2021)Pepe, Cristiani, Rebolo, Santos, Dekker, Cabral,
  Di~Marcantonio, Figueira, Lo~Curto, Lovis, Mayor, Mégevand, Molaro, Riva,
  Zapatero~Osorio, Amate, Manescau, Pasquini, Zerbi, Adibekyan, Abreu,
  Affolter, Alibert, Aliverti, Allart, Allende~Prieto, Álvarez, Alves, Avila,
  Baldini, Bandy, Barros, Benz, Bianco, Borsa, Bourrier, Bouchy, Broeg,
  Calderone, Cirami, Coelho, Conconi, Coretti, Cumani, Cupani, D'Odorico,
  Damasso, Deiries, Delabre, Demangeon, Dumusque, Ehrenreich, Faria, Fragoso,
  Genolet, Genoni, Génova~Santos, González~Hernández, Hughes, Iwert, Kerber,
  Knudstrup, Landoni, Lavie, Lillo-Box, Lizon, Maire, Martins, Mehner, Micela,
  Modigliani, Monteiro, Monteiro, Moschetti, Murphy, Nunes, Oggioni, Oliveira,
  Oshagh, Pallé, Pariani, Poretti, Rasilla, Rebordão, Redaelli,
  Santana~Tschudi, Santin, Santos, Ségransan, Schmidt, Segovia, Sosnowska,
  Sozzetti, Sousa, Spanò, Suárez~Mascareño, Tabernero, Tenegi, Udry, \&
  Zanutta}]{pepe_espresso_2021}
Pepe, F., Cristiani, S., Rebolo, R., {et~al.} 2021, Astronomy and Astrophysics,
  645, A96, aDS Bibcode: 2021A\&A...645A..96P

\bibitem[{Pepper {et~al.}(2007)Pepper, Pogge, DePoy, Marshall, Stanek, Stutz,
  Poindexter, Siverd, O'Brien, Trueblood, \&
  Trueblood}]{pepper_kilodegree_2007}
Pepper, J., Pogge, R.~W., DePoy, D.~L., {et~al.} 2007, Publications of the
  Astronomical Society of the Pacific, 119, 923, aDS Bibcode:
  2007PASP..119..923P

\bibitem[{Perez \& Granger(2007)}]{perez_ipython_2007}
Perez, F. \& Granger, B.~E. 2007, Computing in Science \& Engineering, 9, 21,
  conference Name: Computing in Science \& Engineering

\bibitem[{Pino {et~al.}(2022)Pino, Brogi, Désert, Nascimbeni, Bonomo,
  Rauscher, Basilicata, Biazzo, Bignamini, Borsa, Claudi, Covino, Di~Mauro,
  Guilluy, Maggio, Malavolta, Micela, Molinari, Molinaro, Montalto, Nardiello,
  Pedani, Piotto, Poretti, Rainer, Scandariato, Sicilia, \&
  Sozzetti}]{pino_gaps_2022}
Pino, L., Brogi, M., Désert, J.~M., {et~al.} 2022, The {GAPS} {Programme} at
  {TNG}. {XLI}. {The} climate of {KELT}-9b revealed with a new approach to high
  spectral resolution phase curves, Tech. rep., publication Title: arXiv
  e-prints ADS Bibcode: 2022arXiv220911735P Type: article

\bibitem[{Pollacco {et~al.}(2006)Pollacco, Skillen, Collier~Cameron, Christian,
  Hellier, Irwin, Lister, Street, West, Anderson, Clarkson, Deeg, Enoch, Evans,
  Fitzsimmons, Haswell, Hodgkin, Horne, Kane, Keenan, Maxted, Norton, Osborne,
  Parley, Ryans, Smalley, Wheatley, \& Wilson}]{pollacco_wasp_2006}
Pollacco, D.~L., Skillen, I., Collier~Cameron, A., {et~al.} 2006, Publications
  of the Astronomical Society of the Pacific, 118, 1407, aDS Bibcode:
  2006PASP..118.1407P

\bibitem[{Pont {et~al.}(2008)Pont, Knutson, Gilliland, Moutou, \&
  Charbonneau}]{pont_detection_2008}
Pont, F., Knutson, H., Gilliland, R.~L., Moutou, C., \& Charbonneau, D. 2008,
  Monthly Notices of the Royal Astronomical Society, 385, 109, aDS Bibcode:
  2008MNRAS.385..109P

\bibitem[{Prinoth {et~al.}(2022)Prinoth, Hoeijmakers, Kitzmann, Sandvik,
  Seidel, Lendl, Borsato, Thorsbro, Anderson, Barrado, Kravchenko, Allart,
  Bourrier, Cegla, Ehrenreich, Fisher, Lovis, Guzmán-Mesa, Grimm, Hooton,
  Morris, Oreshenko, Pino, \& Heng}]{prinoth_titanium_2022}
Prinoth, B., Hoeijmakers, H.~J., Kitzmann, D., {et~al.} 2022, Nature Astronomy,
  6, 449, number: 4 Publisher: Nature Publishing Group

\bibitem[{Prinoth {et~al.}(2023)Prinoth, Hoeijmakers, Pelletier, Kitzmann,
  Morris, Seifahrt, Kasper, Korhonen, Burheim, Bean, Benneke, Borsato, Brady,
  Grimm, Luque, Stürmer, \& Thorsbro}]{prinoth_time-resolved_2023}
Prinoth, B., Hoeijmakers, H.~J., Pelletier, S., {et~al.} 2023, Astronomy and
  Astrophysics, 678, A182, aDS Bibcode: 2023A\&A...678A.182P

\bibitem[{Rackham {et~al.}(2018)Rackham, Apai, \&
  Giampapa}]{rackham_transit_2018}
Rackham, B.~V., Apai, D., \& Giampapa, M.~S. 2018, The Astrophysical Journal,
  853, 122, aDS Bibcode: 2018ApJ...853..122R

\bibitem[{Rauer {et~al.}(2011)Rauer, Gebauer, Paris, Cabrera, Godolt, Grenfell,
  Belu, Selsis, Hedelt, \& Schreier}]{rauer_potential_2011}
Rauer, H., Gebauer, S., Paris, P.~V., {et~al.} 2011, Astronomy and
  Astrophysics, 529, A8, aDS Bibcode: 2011A\&A...529A...8R

\bibitem[{Redfield {et~al.}(2008)Redfield, Endl, Cochran, \&
  Koesterke}]{redfield_sodium_2008}
Redfield, S., Endl, M., Cochran, W.~D., \& Koesterke, L. 2008, The
  Astrophysical Journal, 673, L87, aDS Bibcode: 2008ApJ...673L..87R

\bibitem[{Ricker {et~al.}(2015)Ricker, Winn, Vanderspek, Latham, Bakos, Bean,
  Berta-Thompson, Brown, Buchhave, Butler, Butler, Chaplin, Charbonneau,
  Christensen-Dalsgaard, Clampin, Deming, Doty, De~Lee, Dressing, Dunham, Endl,
  Fressin, Ge, Henning, Holman, Howard, Ida, Jenkins, Jernigan, Johnson,
  Kaltenegger, Kawai, Kjeldsen, Laughlin, Levine, Lin, Lissauer, MacQueen,
  Marcy, McCullough, Morton, Narita, Paegert, Palle, Pepe, Pepper, Quirrenbach,
  Rinehart, Sasselov, Sato, Seager, Sozzetti, Stassun, Sullivan, Szentgyorgyi,
  Torres, Udry, \& Villasenor}]{ricker_transiting_2015}
Ricker, G.~R., Winn, J.~N., Vanderspek, R., {et~al.} 2015, Journal of
  Astronomical Telescopes, Instruments, and Systems, 1, 014003, aDS Bibcode:
  2015JATIS...1a4003R

\bibitem[{Ridden-Harper {et~al.}(2023)Ridden-Harper, de~Mooij, Jayawardhana,
  Gibson, Karjalainen, \& Karjalainen}]{ridden-harper_high-resolution_2023}
Ridden-Harper, A., de~Mooij, E., Jayawardhana, R., {et~al.} 2023, The
  Astronomical Journal, 165, 211, aDS Bibcode: 2023AJ....165..211R

\bibitem[{Rothman {et~al.}(2010)Rothman, Gordon, Barber, Dothe, Gamache,
  Goldman, Perevalov, Tashkun, \& Tennyson}]{rothman_hitemp_2010}
Rothman, L.~S., Gordon, I.~E., Barber, R.~J., {et~al.} 2010, Journal of
  Quantitative Spectroscopy and Radiative Transfer, 111, 2139

\bibitem[{Rustamkulov {et~al.}(2023)Rustamkulov, Sing, Mukherjee, May, Line,
  Schlawin, Stevenson, Piaulet, Moran, Carter, Kirk, Batalha, Lothringer,
  Lopez-Morales, Bean, Espinoza, Batalha, \& {JWST Transiting Exoplanet
  Community ERS Team}}]{rustamkulov_jwst_2023}
Rustamkulov, Z., Sing, D., Mukherjee, S., {et~al.} 2023, Conference Name:
  American Astronomical Society Meeting Abstracts, 55, 124.01, aDS Bibcode:
  2023AAS...24112401R

\bibitem[{Sabotta {et~al.}(2021)Sabotta, Schlecker, Chaturvedi, Guenther,
  Rodríguez, Sánchez, Caballero, Shan, Reffert, Ribas, Reiners, Hatzes,
  Amado, Klahr, Morales, Quirrenbach, Henning, Dreizler, Pallé, Perger,
  Azzaro, Jeffers, Kaminski, Kürster, Lafarga, Montes, Passegger, \&
  Zechmeister}]{sabotta_carmenes_2021}
Sabotta, S., Schlecker, M., Chaturvedi, P., {et~al.} 2021, Astronomy \&
  Astrophysics, 653, A114, publisher: EDP Sciences

\bibitem[{Scandariato {et~al.}(2023)Scandariato, Borsa, Bonomo, Gaudi, Henning,
  Ilyin, Johnson, Malavolta, Mallonn, Molaverdikhani, Nascimbeni, Patience,
  Pino, Poppenhaeger, Schlawin, Shkolnik, Sicilia, Sozzetti, Strassmeier,
  Veillet, Wang, \& Yan}]{scandariato_pepsi_2023}
Scandariato, G., Borsa, F., Bonomo, A.~S., {et~al.} 2023, Astronomy and
  Astrophysics, 674, A58, aDS Bibcode: 2023A\&A...674A..58S

\bibitem[{Seager \& Sasselov(2000)}]{seager_theoretical_2000}
Seager, S. \& Sasselov, D.~D. 2000, The Astrophysical Journal, 537, 916, aDS
  Bibcode: 2000ApJ...537..916S

\bibitem[{Seidel {et~al.}(2020{\natexlab{a}})Seidel, Ehrenreich, Pino,
  Bourrier, Lavie, Allart, Wyttenbach, \& Lovis}]{seidel_wind_2020}
Seidel, J.~V., Ehrenreich, D., Pino, L., {et~al.} 2020{\natexlab{a}}, Astronomy
  \& Astrophysics, 633, A86, publisher: EDP Sciences

\bibitem[{Seidel {et~al.}(2020{\natexlab{b}})Seidel, Lendl, Bourrier,
  Ehrenreich, Allart, Sousa, Cegla, Bonfils, Conod, Grandjean, Wyttenbach,
  Astudillo-Defru, Bayliss, Heng, Lavie, Lovis, Melo, Pepe, Ségransan, \&
  Udry}]{seidel_hot_2020}
Seidel, J.~V., Lendl, M., Bourrier, V., {et~al.} 2020{\natexlab{b}}, Astronomy
  and Astrophysics, 643, A45, aDS Bibcode: 2020A\&A...643A..45S

\bibitem[{Seifahrt {et~al.}(2020)Seifahrt, Bean, Stürmer, Kasper, Gers,
  Schwab, Zechmeister, Stefánsson, Montet, Dos~Santos, Peck, White, \&
  Tapia}]{seifahrt_-sky_2020}
Seifahrt, A., Bean, J.~L., Stürmer, J., {et~al.} 2020, Conference Name:
  Ground-based and Airborne Instrumentation for Astronomy VIII, 11447, 114471F,
  conference Name: Ground-based and Airborne Instrumentation for Astronomy VIII
  ADS Bibcode: 2020SPIE11447E..1FS

\bibitem[{Sing {et~al.}(2011)Sing, Pont, Aigrain, Charbonneau, Désert, Gibson,
  Gilliland, Hayek, Henry, Knutson, Lecavelier Des~Etangs, Mazeh, \&
  Shporer}]{sing_hubble_2011}
Sing, D.~K., Pont, F., Aigrain, S., {et~al.} 2011, Monthly Notices of the Royal
  Astronomical Society, 416, 1443, aDS Bibcode: 2011MNRAS.416.1443S

\bibitem[{Sing {et~al.}(2008)Sing, Vidal-Madjar, Désert, Lecavelier~des
  Etangs, \& Ballester}]{sing_hubble_2008}
Sing, D.~K., Vidal-Madjar, A., Désert, J.~M., Lecavelier~des Etangs, A., \&
  Ballester, G. 2008, The Astrophysical Journal, 686, 658, aDS Bibcode:
  2008ApJ...686..658S

\bibitem[{Smette {et~al.}(2015)Smette, Sana, Noll, Horst, Kausch, Kimeswenger,
  Barden, Szyszka, Jones, Gallenne, Vinther, Ballester, \&
  Taylor}]{smette_molecfit_2015}
Smette, A., Sana, H., Noll, S., {et~al.} 2015, Astronomy \& Astrophysics, 576,
  A77, publisher: EDP Sciences

\bibitem[{Snellen {et~al.}(2015)Snellen, Kok, Birkby, Brandl, Brogi, Keller,
  Kenworthy, Schwarz, \& Stuik}]{snellen_combining_2015}
Snellen, I., Kok, R.~d., Birkby, J.~L., {et~al.} 2015, Astronomy \&
  Astrophysics, 576, A59, publisher: EDP Sciences

\bibitem[{Snellen {et~al.}(2008)Snellen, Albrecht, de~Mooij, \&
  Le~Poole}]{snellen_ground-based_2008}
Snellen, I. A.~G., Albrecht, S., de~Mooij, E. J.~W., \& Le~Poole, R.~S. 2008,
  Astronomy and Astrophysics, 487, 357, aDS Bibcode: 2008A\&A...487..357S

\bibitem[{Snellen {et~al.}(2010)Snellen, de~Kok, de~Mooij, \&
  Albrecht}]{snellen_orbital_2010}
Snellen, I. A.~G., de~Kok, R.~J., de~Mooij, E. J.~W., \& Albrecht, S. 2010,
  Nature, 465, 1049

\bibitem[{Spyratos {et~al.}(2023)Spyratos, Nikolov, Constantinou, Southworth,
  Madhusudhan, Sedaghati, Ehrenreich, \& Mancini}]{spyratos_precise_2023}
Spyratos, P., Nikolov, N.~K., Constantinou, S., {et~al.} 2023, Monthly Notices
  of the Royal Astronomical Society, 521, 2163, aDS Bibcode:
  2023MNRAS.521.2163S

\bibitem[{Stassun {et~al.}(2019)Stassun, Oelkers, Paegert, Torres, Pepper, Lee,
  Collins, Latham, Muirhead, Chittidi, Rojas-Ayala, Fleming, Rose, Tenenbaum,
  Ting, Kane, Barclay, Bean, Brassuer, Charbonneau, Ge, Lissauer, Mann, McLean,
  Mullally, Narita, Plavchan, Ricker, Sasselov, Seager, Sharma, Shiao,
  Sozzetti, Stello, Vanderspek, Wallace, \& Winn}]{stassun_revised_2019}
Stassun, K.~G., Oelkers, R.~J., Paegert, M., {et~al.} 2019, The Astronomical
  Journal, 158, 138, publisher: The American Astronomical Society

\bibitem[{Stevenson(2016)}]{stevenson_quantifying_2016}
Stevenson, K.~B. 2016, The Astrophysical Journal, 817, L16, aDS Bibcode:
  2016ApJ...817L..16S

\bibitem[{Swain {et~al.}(2008)Swain, Vasisht, \& Tinetti}]{swain_presence_2008}
Swain, M.~R., Vasisht, G., \& Tinetti, G. 2008, Nature, 452, 329, number: 7185
  Publisher: Nature Publishing Group

\bibitem[{Sánchez-López {et~al.}(2019)Sánchez-López, Alonso-Floriano,
  López-Puertas, Snellen, Funke, Nagel, Bauer, Amado, Caballero, Czesla,
  Nortmann, Pallé, Salz, Reiners, Ribas, Quirrenbach, Anglada-Escudé, Béjar,
  Casasayas-Barris, Galadí-Enríquez, Guenther, Henning, Kaminski, Kürster,
  Lampón, Lara, Montes, Morales, Stangret, Tal-Or, Sanz-Forcada, Schmitt,
  Osorio, \& Zechmeister}]{sanchez-lopez_water_2019}
Sánchez-López, A., Alonso-Floriano, F.~J., López-Puertas, M., {et~al.} 2019,
  Astronomy \& Astrophysics, 630, A53, publisher: EDP Sciences

\bibitem[{Talens {et~al.}(2017)Talens, Albrecht, Spronck, Lesage, Otten, Stuik,
  Van~Eylen, Van~Winckel, Pollacco, McCormac, Grundahl, Andersen, Antoci, \&
  Snellen}]{talens_mascara-1_2017}
Talens, G. J.~J., Albrecht, S., Spronck, J. F.~P., {et~al.} 2017, Astronomy \&
  Astrophysics, 606, A73, arXiv:1707.04262 [astro-ph]

\bibitem[{Tamuz {et~al.}(2005)Tamuz, Mazeh, \& Zucker}]{tamuz_correcting_2005}
Tamuz, O., Mazeh, T., \& Zucker, S. 2005, Monthly Notices of the Royal
  Astronomical Society, 356, 1466

\bibitem[{team(2023)}]{team_pandas-devpandas_2023}
team, T. p.~d. 2023, pandas-dev/pandas: {Pandas} (v2.0.3)

\bibitem[{Van~Eylen {et~al.}(2019)Van~Eylen, Albrecht, Huang, MacDonald,
  Dawson, Cai, Foreman-Mackey, Lundkvist, Silva~Aguirre, Snellen, \&
  Winn}]{van_eylen_orbital_2019}
Van~Eylen, V., Albrecht, S., Huang, X., {et~al.} 2019, The Astronomical
  Journal, 157, 61, aDS Bibcode: 2019AJ....157...61V

\bibitem[{Vidal-Madjar {et~al.}(2004)Vidal-Madjar, Désert, Lecavelier~des
  Etangs, Hébrard, Ballester, Ehrenreich, Ferlet, McConnell, Mayor, \&
  Parkinson}]{vidal-madjar_detection_2004}
Vidal-Madjar, A., Désert, J.~M., Lecavelier~des Etangs, A., {et~al.} 2004, The
  Astrophysical Journal, 604, L69, aDS Bibcode: 2004ApJ...604L..69V

\bibitem[{Virtanen {et~al.}(2020)Virtanen, Gommers, Oliphant, Haberland, Reddy,
  Cournapeau, Burovski, Peterson, Weckesser, Bright, van~der Walt, Brett,
  Wilson, Millman, Mayorov, Nelson, Jones, Kern, Larson, Carey, Polat, Feng,
  Moore, VanderPlas, Laxalde, Perktold, Cimrman, Henriksen, Quintero, Harris,
  Archibald, Ribeiro, Pedregosa, \& van Mulbregt}]{virtanen_scipy_2020}
Virtanen, P., Gommers, R., Oliphant, T.~E., {et~al.} 2020, Nature Methods, 17,
  261, number: 3 Publisher: Nature Publishing Group

\bibitem[{Wakeford {et~al.}(2019)Wakeford, Lewis, Fowler, Bruno, Wilson, Moran,
  Valenti, Batalha, Filippazzo, Bourrier, Hörst, Lederer, \&
  de~Wit}]{wakeford_disentangling_2019}
Wakeford, H.~R., Lewis, N.~K., Fowler, J., {et~al.} 2019, The Astronomical
  Journal, 157, 11, aDS Bibcode: 2019AJ....157...11W

\bibitem[{Waskom {et~al.}(2017)Waskom, Botvinnik, O'Kane, Hobson, Lukauskas,
  Gemperline, Augspurger, Halchenko, Cole, Warmenhoven, de~Ruiter, Pye, Hoyer,
  Vanderplas, Villalba, Kunter, Quintero, Bachant, Martin, Meyer, Miles, Ram,
  Yarkoni, Williams, Evans, Fitzgerald, {Brian}, Fonnesbeck, Lee, \&
  Qalieh}]{waskom_mwaskomseaborn_2017}
Waskom, M., Botvinnik, O., O'Kane, D., {et~al.} 2017, Zenodo, aDS Bibcode:
  2017zndo....883859W

\bibitem[{Way \& Del~Genio(2020)}]{way_venusian_2020}
Way, M.~J. \& Del~Genio, A.~D. 2020, Journal of Geophysical Research: Planets,
  125, e2019JE006276, \_eprint:
  https://onlinelibrary.wiley.com/doi/pdf/10.1029/2019JE006276

\bibitem[{Wolf {et~al.}(2019)Wolf, Kopparapu, Airapetian, Fauchez, Guzewich,
  Kane, Pidhorodetska, Way, Abbot, Checlair, Davis, Del~Genio, Dong, Eggl,
  Fleming, Fujii, Haghighipour, Heavens, Henning, Kiang, Lopez-Morales,
  Lustig-Yaeger, Meadows, Reinhard, Rugheimer, Schwieterman, Shields, Sohl,
  Turbet, \& Wordsworth}]{wolf_importance_2019}
Wolf, E.~T., Kopparapu, R., Airapetian, V., {et~al.} 2019, The {Importance} of
  {3D} {General} {Circulation} {Models} for {Characterizing} the {Climate} and
  {Habitability} of {Terrestrial} {Extrasolar} {Planets}, arXiv:1903.05012
  [astro-ph]

\bibitem[{Wunderlich {et~al.}(2019)Wunderlich, Godolt, Grenfell, Städt, Smith,
  Gebauer, Schreier, Hedelt, \& Rauer}]{wunderlich_detectability_2019}
Wunderlich, F., Godolt, M., Grenfell, J.~L., {et~al.} 2019, Astronomy \&
  Astrophysics, 624, A49, publisher: EDP Sciences

\bibitem[{Yan {et~al.}(2023)Yan, Nortmann, Reiners, Piskunov, Hatzes, Seemann,
  Shulyak, Lavail, Rains, Cont, Rengel, Lesjak, Nagel, Kochukhov, Czesla,
  Boldt-Christmas, Heiter, Smoker, Rodler, Bristow, Dorn, Jung, Marquart, \&
  Stempels}]{yan_crires_2023}
Yan, F., Nortmann, L., Reiners, A., {et~al.} 2023, Astronomy and Astrophysics,
  672, A107, aDS Bibcode: 2023A\&A...672A.107Y

\bibitem[{Zhou {et~al.}(2022)Zhou, Ma, Wang, \& Zhu}]{zhou_hubble_2022}
Zhou, L., Ma, B., Wang, Y., \& Zhu, Y. 2022, The Astronomical Journal, 164,
  203, aDS Bibcode: 2022AJ....164..203Z

\end{thebibliography}

\end{document}